\documentclass[superscriptaddress,aps,preprintnumbers,amsmath,amssymb,prd,nofootinbib,preprint,a4paper]{revtex4-1}

\usepackage{graphicx}
\usepackage{epstopdf}
\usepackage{dcolumn}
\usepackage{bm}
\usepackage{hyperref}
\usepackage{color}
\usepackage{setspace}

\begin{document}


\def\a{\alpha}
\def\b{\beta}
\def\c{\varepsilon}
\def\d{\delta}
\def\e{\epsilon}
\def\f{\phi}
\def\g{\gamma}
\def\h{\theta}
\def\k{\kappa}
\def\l{\lambda}
\def\m{\mu}
\def\n{\nu}
\def\p{\psi}
\def\q{\partial}
\def\r{\rho}
\def\s{\sigma}
\def\t{\tau}
\def\u{\upsilon}
\def\v{\varphi}
\def\w{\omega}
\def\x{\xi}
\def\y{\eta}
\def\z{\zeta}
\def\D{\Delta}
\def\G{\Gamma}
\def\H{\Theta}
\def\L{\Lambda}
\def\F{\Phi}
\def\P{\Psi}
\def\S{\Sigma}

\def\o{\over}
\def\beq{\begin{eqnarray}}
\def\eeq{\end{eqnarray}}
\newcommand{\gsim}{ \mathop{}_{\textstyle \sim}^{\textstyle >} }
\newcommand{\lsim}{ \mathop{}_{\textstyle \sim}^{\textstyle <} }
\newcommand{\vev}[1]{ \left\langle {#1} \right\rangle }
\newcommand{\bra}[1]{ \langle {#1} | }
\newcommand{\ket}[1]{ | {#1} \rangle }
\newcommand{\EV}{ {\rm eV} }
\newcommand{\KEV}{ {\rm keV} }
\newcommand{\MEV}{ {\rm MeV} }
\newcommand{\GEV}{ {\rm GeV} }
\newcommand{\TEV}{ {\rm TeV} }
\newcommand{\1}{\mbox{1}\hspace{-0.25em}\mbox{l}}
\newcommand{\headline}[1]{\noindent{\bf #1}}
\def\diag{\mathop{\rm diag}\nolimits}
\def\Spin{\mathop{\rm Spin}}
\def\SO{\mathop{\rm SO}}
\def\O{\mathop{\rm O}}
\def\SU{\mathop{\rm SU}}
\def\U{\mathop{\rm U}}
\def\Sp{\mathop{\rm Sp}}
\def\SL{\mathop{\rm SL}}
\def\tr{\mathop{\rm tr}}
\def\mpl{M_{PL}}

\def\IJMP{Int.~J.~Mod.~Phys. }
\def\MPL{Mod.~Phys.~Lett. }
\def\NP{Nucl.~Phys. }
\def\PL{Phys.~Lett. }
\def\PR{Phys.~Rev. }
\def\PRL{Phys.~Rev.~Lett. }
\def\PTP{Prog.~Theor.~Phys. }
\def\ZP{Z.~Phys. }

\def\dd{\mathrm{d}}
\def\ff{\mathrm{f}}
\def\BH{{\rm BH}}
\def\inf{{\rm inf}}
\def\ev{{\rm evap}}
\def\eq{{\rm eq}}
\def\SM{{\rm sm}}
\def\Mpl{M_{\rm Pl}}
\def\GeV{{\rm GeV}}
\newcommand{\Red}[1]{\textcolor{red}{#1}}

%


\title{Revisiting $R$-invariant Direct Gauge Mediation}
\author{Cheng-Wei Chiang}
\affiliation{Center for Mathematics and Theoretical Physics and Department of Physics, National Central University, Taoyuan, Taiwan 32001, R.O.C.}
\affiliation{Institute of Physics, Academia Sinica, Taipei, Taiwan 11529, R.O.C.}
\affiliation{Physics Division, National Center for Theoretical Sciences, Hsinchu, Taiwan 30013, R.O.C.}
\affiliation{Kavli IPMU (WPI), UTIAS, University of Tokyo, Kashiwa, Chiba 277-8583, Japan}
\author{Keisuke~Harigaya}
\affiliation{Department of Physics, University of California, Berkeley, California 94720, USA}
\affiliation{Theoretical Physics Group, Lawrence Berkeley National Laboratory, Berkeley, California 94720, USA}
\affiliation{ICRR, University of Tokyo, Kashiwa, Chiba 277-8582, Japan}
\author{Masahiro~Ibe}
\affiliation{Kavli IPMU (WPI), UTIAS, University of Tokyo, Kashiwa, Chiba 277-8583, Japan}
\affiliation{ICRR, University of Tokyo, Kashiwa, Chiba 277-8582, Japan}
\author{Tsutomu T. Yanagida}
\affiliation{Kavli IPMU (WPI), UTIAS, University of Tokyo, Kashiwa, Chiba 277-8583, Japan}

\begin{abstract}
We revisit a special model of gauge mediated supersymmetry breaking, the ``$R$-invariant direct gauge mediation.''
We pay particular attention to whether the model is consistent with the minimal model of the $\mu$-term,
{\it i.e.}, a simple mass term of the Higgs doublets in the superpotential.
Although the incompatibility is highlighted in view of the current experimental constraints on the superparticle masses and the observed Higgs boson mass, the minimal $\mu$-term can be consistent with the $R$-invariant gauge mediation model
via a careful choice of model parameters.
We derive an upper limit on the gluino mass from the observed Higgs boson mass.
We also discuss whether the model can explain the $3\sigma$ excess of the $Z+$jets$+E_T^{\rm miss}$ events reported by the ATLAS Collaboration.
\end{abstract}

\date{\today}
\maketitle
\preprint{IPMU15-0180}

\section{Introduction}
The model of gauge mediated supersymmetry (SUSY) breaking \,\cite{Dine:1981za,Dine:1981gu} is the most attractive candidate for phenomenologically successful minimal supersymmetric standard model (MSSM).  In this case, soft SUSY breaking is mediated via the MSSM gauge interactions and, thus, the model is free from the infamous SUSY flavor changing neutral current problem.

One of the drawbacks of gauge mediation models is their somewhat cumbersome structure.
In particular, careful model building is required to connect 
messenger fields to a SUSY breaking sector without destabilizing the SUSY breaking vacuum in the SUSY breaking sector.
In fact, naive couplings between the SUSY breaking sector and messenger fields often lead
to meta-stability of the SUSY breaking vacuum.
In those models, the thermal history of the Universe and/or the masses of messenger fields
are severely constrained\,\cite{Hisano:2008sy}.

Among various safe scheme to connect messengers to the SUSY breaking sector,
the model developed in Refs.~\cite{Izawa:1997gs,Nomura:1997uu} is highly successful.
In particular, the SUSY breaking vacuum is not destabilized, and hence the model
is durable even when the reheating temperature of the Universe is very high.
The stability of the SUSY breaking vacuum is achieved through
$R$-symmetry, which is the origin of the name of the model, ``$R$-invariant direct gauge mediation.''%
\footnote{For simple embedding of the model into a dynamical SUSY breaking model
with a radiative $R$-symmetry breaking, see Ref.~\cite{Ibe:2010ym}.}

In this paper, we revisit the $R$-invariant direct gauge mediation model by 
paying particular attention to the consistency of the model with the minimal model
that addresses the origin of the $\mu$-term. 
Here the minimal model of the $\mu$-term means one with a simple mass term for the
Higgs doublets in the superpotential, which leads to
a vanishing $B$-term at the messenger scale.
As pointed out in Ref.~\cite{Ibe:2012dd}, it is difficult for the minimal model of the $\mu$-term 
to be compatible with the $R$-invariant direct gauge mediation,
for the model predicts rather suppressed gaugino masses compared with scalar masses.
As we will see in this paper, the minimal $\mu$-term can be consistent with the $R$-invariant gauge mediation model
through a careful choice of model parameters, although the incompatibility is highlighted in view of the current 
experimental constraints on superparticle masses and the observed Higgs boson mass.

After discussing the compatibility of the  $R$-invariant direct gauge mediation model with the minimal $\mu$-term,
we derive an upper bound on the gluino mass from the observed Higgs boson mass by exploiting 
the predicted  ratio between the gluino and the stop masses in the $R$-invariant gauge mediation model.
As a result of the upper bound, we find that a large portion of parameter space can 
be tested by the LHC Run-II with an integrated luminosity of $300\,$fb$^{-1}$, unless model parameters are highly optimized to obtain a large gluino mass.

We also discuss whether the $R$-invariant direct gauge mediation model can explain the
$3\sigma$ excess of $Z+$jets$+E_T^{\rm miss}$ events reported by the ATLAS Collaboration\,\cite{Aad:2015wqa}. 
We seek a spectrum similar to the one in Ref.~\cite{Lu:2015wwa} where the gluino mainly decays into a 
gluon and a Higgsino via one-loop corrections.
With such a spectrum, the excess can be explained while evading all the other constraints from SUSY searches at the LHC.

The paper is organized as follows.
In section\,\ref{sec:model}, we discuss the consistency between the $R$-invariant direct gauge mediation
model and the minimal model of the $\mu$-term.
In section\,\ref{sec:gluino}, we derive an upper bound on the gluino mass from the observed Higgs boson mass.
In section\,\ref{sec:signal}, we discuss whether the $R$-invariant direct gauge mediation
can explain the signal reported by the ATLAS Collaboration.
The last section is devoted to the summary of our discussions.

\section{$B\mu$--problem in $R$-invariant Direct Gauge Mediation Model}\label{sec:model}

\subsection{$R$-invariant Direct Gauge Mediation Model}
\label{sec:RDGMB}
We first review the minimal $R$-invariant direct gauge mediation model
constructed in Ref.~\cite{Izawa:1997gs, Nomura:1997uu} (see also Ref.~\cite{Cheung:2007es}). 			
This model introduces $N_M$ sets of messenger fields, $\Psi_i$, $\bar \Psi_i$, 
$\Psi_i'$ and $\bar \Psi_i'$, 
which are respectively ${\bf 5}$,  $\bar {\bf 5}$, ${\bf 5}$ and  $\bar {\bf 5}$  representations of the $SU(5)$ gauge group of the grand unified theory (GUT).
The index $i = 1,\cdots,N_M$ labels each set of messengers.

Messengers of each set directly couple to a supersymmetry breaking gauge singlet field $S$
in the superpotential,
\begin{eqnarray}
\label{eq:SUSY-MESS}
W =W_{\rm SUSY} + k S \Psi_i \bar\Psi_i + M_{\Psi} \Psi_i \bar{\Psi}'_i  + M_{\bar\Psi} \bar\Psi_i {\Psi}'_i \ ,
\end{eqnarray}
where $k$ denotes a coupling constant and $M_{\Psi,\bar\Psi}$ are mass parameters.
$W_{\rm SUSY}$ encapsulates a dynamical SUSY breaking sector such as those in Ref.~\cite{Izawa:1996pk,Intriligator:1996pu,Intriligator:2006dd} whose effective theory is simply given by
\begin{eqnarray}
W_{\rm SUSY} \simeq \Lambda^2 S ~,
\end{eqnarray}
where $\Lambda$ denotes the associated dynamical scale.  Due to the linear term of $S$ in the superpotential, the SUSY breaking field obtains a non-vanishing $F$-term expectation value.
It should be noted that the form of the superpotential in Eq.\,(\ref{eq:SUSY-MESS}) is protected by
an $R$-symmetry with the charge assignments 
$S(2)$, 
$\Psi_i(0)$, 
$\bar \Psi_i(0)$, 
$\Psi_i'(2)$ 
and $\bar \Psi_i'(2)$,%
\footnote{The charges are assigned up to $U(1)$ messenger symmetries which are eventually broken by 
mixing with MSSM fields.}
which gives the origin of the name of the $R$-invariant direct gauge mediation (see also appendix\,\ref{sec:Rcharge}).
Due to this peculiar form of the superpotential, the SUSY breaking vacuum is not destabilized by the 
couplings to the messenger fields.

It should be emphasized that the $R$-symmetry needs to be broken spontaneously to generate non-vanishing 
MSSM gaugino masses. 
Such spontaneous $R$-symmetry breaking can be achieved, for example, 
through a simple extension of the dynamical SUSY breaking model\,\cite{Izawa:1996pk,Intriligator:1996pu} with an extra $U(1)$ gauge interaction\,\cite{Ibe:2010ym}.
See also Refs.~\cite{Shih:2007av,Intriligator:2007py,Giveon:2008ne,Evans:2011pz,Curtin:2012yu} 
for radiative $R$-symmetry breaking in more generic models.%
\footnote{It is also possible to construct O'Raifeartaigh models where
spontaneous SUSY and $R$-symmetry breakings are achieved at {\it tree level} 
\,\cite{Carpenter:2008wi,Sun:2008va,Komargodski:2009jf}.}
Altogether, we postulate that the SUSY breaking field $S$ obtains its expectation value,
\begin{eqnarray}
   \vev{S(x,\theta)}  = S_0 + F\,\theta^2\ ,
\end{eqnarray}
where $S_0$ denotes the vacuum expectation value of the $A$-term of $S$ 
and $\theta$ is the fermionic coordinate of the superspace.

Due to the stability of the SUSY breaking vacuum, the model is viable even when the reheating temperature of the 
Universe is very high. 
Thus, the model is consistent with thermal leptogenesis with
$T_R \gtrsim 10^9$\,GeV\,\cite{leptogenesis}.
This feature should be compared with other types of direct gauge mediation models
where a SUSY breaking vacuum is destabilized by messenger couplings such as
\begin{eqnarray}
W =W_{\rm SUSY} + k S \Psi_i \Psi_i \ ,
\end{eqnarray}
 (see {\it e.g.}, Ref.~\cite{Ibe:2007ab}).
In such cases, the thermal history of the Universe and/or the masses of the messenger fields
are severely restricted\,\cite{Hisano:2008sy}.%
\footnote{
For phenomenological studies of this class of models after the LHC Run-I
experiment, see {\it e.g.}, Ref.~\cite{Hamaguchi:2014sea}. }

\subsection{Gauge Mediated Mass Spectrum}

\begin{figure}[t]
\begin{center}
\begin{minipage}{.46\linewidth}
  \includegraphics[width=\linewidth]{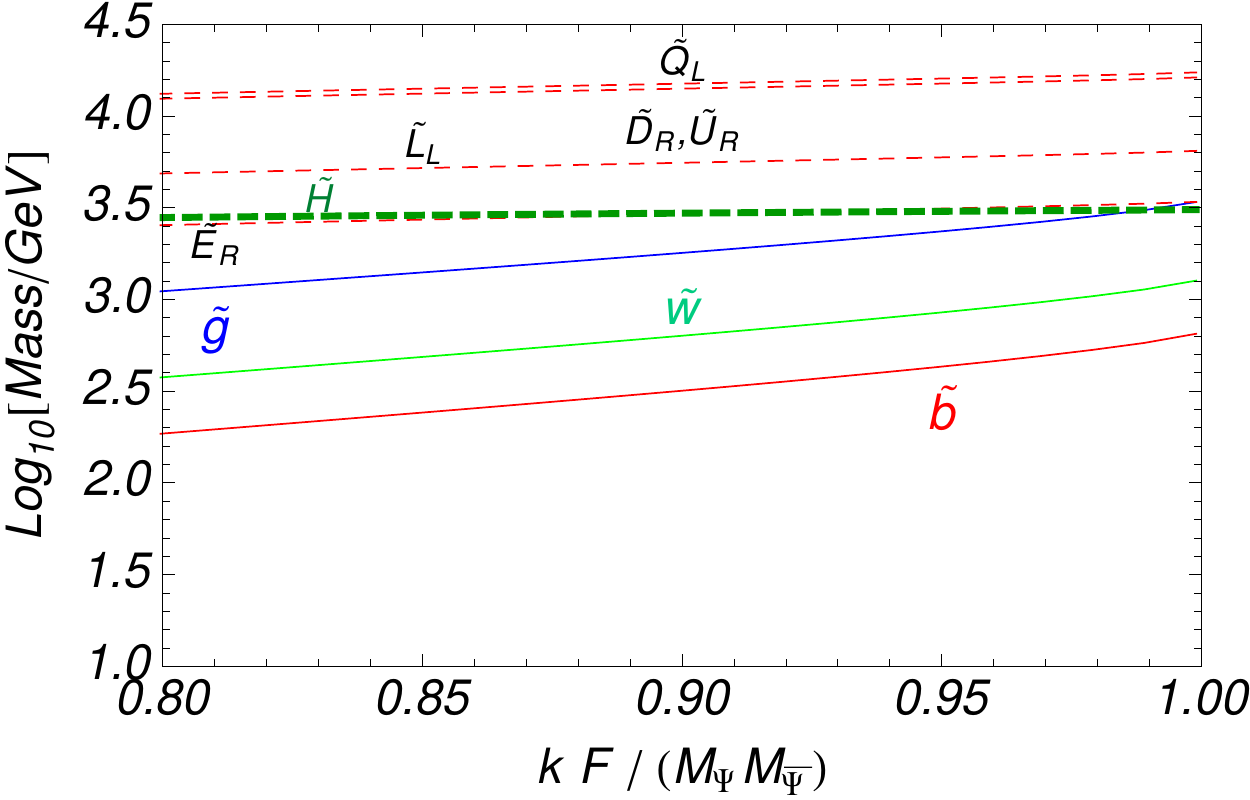}
 \end{minipage}
 \hspace{1cm}
 \begin{minipage}{.46\linewidth}
  \includegraphics[width=\linewidth]{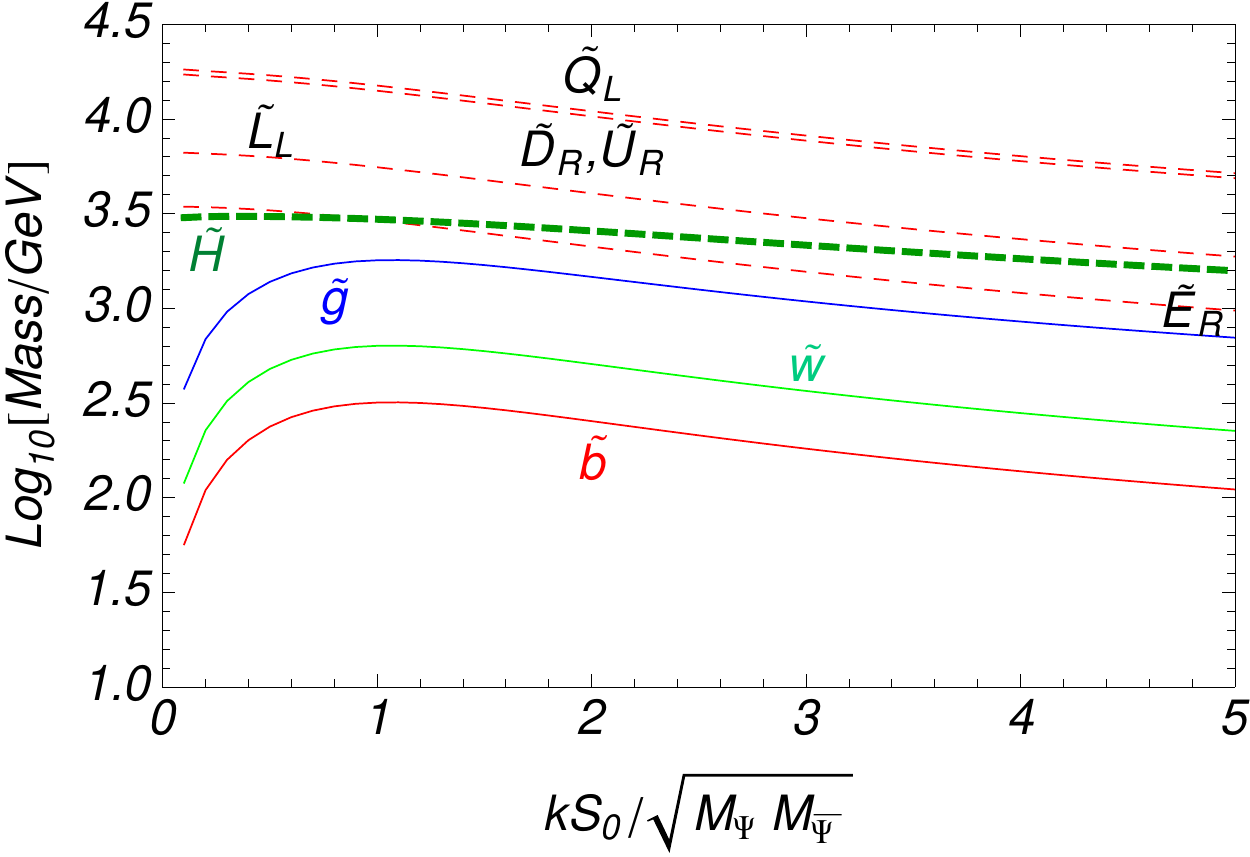}
 \end{minipage}
 \end{center}
\caption{\sl \small
The gauge mediated mass spectrum. 
The curves give the masses of various sparticles. 
In the figure, we take $M_{\Psi} = M_{\bar \Psi} = 2\times 10^6$\,GeV,
$N_M = 1$ and $\tan\beta = 10$.
In the left plot, we take $kS_0/\sqrt{M_{\Psi}M_{\bar \Psi}} = 1$ and show the spectrum
as a function of $kF/(M_{\Psi}M_{\bar \Psi})$.
In the right plot, we take $kF/(M_{\Psi}M_{\bar \Psi}) = 0.9$ and show the spectrum
as a function of $kS_0/\sqrt{M_{\Psi}M_{\bar \Psi}}$.
}
\label{fig:spectrum}
\end{figure}

We now summarize the gauge mediated mass spectrum of MSSM particles.
The most distinctive feature of the MSSM spectrum in the $R$-invariant direct gauge mediation
is that gaugino masses vanish at the one-loop level to the leading order of the SUSY breaking parameter, ${\cal O}(kF/M_{\rm mess})$~\cite{Izawa:1997gs, Nomura:1997uu} and are suppressed by a factor of ${\cal O}(k^2F^2/M_{\rm mess}^4)$ in comparison with those in the conventional gauge mediation.
In the following, we collectively denote the mass scale of the messenger sector by $M_{\rm mess}$.
Scalar masses, on the other hand, appear at the leading order of the SUSY breaking parameter at the 
two-loop level.
Therefore, gauge mediated MSSM gaugino masses, $M_a$ ($a = 1,2,3$), and MSSM scalar masses, $m_{\rm scalar}$, 
are roughly given by
\begin{eqnarray}
\label{eq:gaugino}
M_{a} &\sim&\frac{g_a^2}{16\pi} \frac{k F}{M_{\rm mess}}\times {\cal O}\left(\frac{k^2F^2}{M_{\rm mess}^4}\right)\times (0.1-0.3)\ ,\\ 
\label{eq:scalar}
m_{\rm scalar}   &\sim& \frac{g_a^2}{16\pi} \frac{k F}{M_{\rm mess}} \ ,
\end{eqnarray}
where $g_a$ ($a = 1,2,3$) denote the gauge coupling constants of the MSSM gauge interactions.
A  factor of ${\cal O}(0.1)$ at the end of Eq.\,(\ref{eq:gaugino}) for the gaugino masses results from numerical analyses (see Fig.\,\ref{fig:spectrum} and the following discussions).
As a result, the predicted spectrum is hierarchical between gaugino masses and sfermion masses.

To date, searches for  gluino pair production at the ATLAS and the CMS experiments 
have put severe lower limits on the gluino mass at around $1.4$\,TeV at 95\%\,CL. 
The limits are applicable for cases where the bino either is stable~\cite{Khachatryan:2015vra,Aad:2015iea} or decays 
into a photon and a gravitino inside the detectors
as the next-to-the lightest superparticle (NLSP)~\cite{Khachatryan:2015exa,Aad:2015hea}.
To satisfy this constraint, we infer 
that
\begin{eqnarray}
\label{eq:gauginobound}
\frac{k F}{M_{\rm mess}}\times {\cal O}\left(\frac{k^2F^2}{M_{\rm mess}^4}\right) = 10^{6-7}\,{\rm GeV}\ ,
\end{eqnarray}
so that the gluino is sufficiently heavy (see Eq.\,(\ref{eq:gaugino})).

Due to the hierarchy between the gaugino masses and sfermion masses in Eqs.\,(\ref{eq:gaugino}) and (\ref{eq:scalar}),
the squarks are beyond the reach of the LHC Run-I when the gluino is heavier than $1.4$\,TeV.
On the other hand, it should be noted that the squark masses are bounded from ``above''
by the correlation between the squark masses and the predicted lightest Higgs boson mass in the MSSM.
In fact, unless the ratio of the Higgs vacuum expectation values, $\tan\beta$, is very close to unity, 
the scalar mass (especially the stop mass) should be around $10-100$\,TeV 
so that the lightest Higgs boson mass is consistent with the observed value\,\cite{OYY}, 
$m_h = 125.09\pm 0.21\pm 0.11$\,GeV\,\cite{Aad:2015zhl}
(see discussions in section\,\ref{sec:gluino} for details). 
This requirement roughly leads to
\begin{eqnarray}
\label{eq:scalarbound}
\frac{k F}{M_{\rm mess}} = 10^{6-7}\,{\rm GeV}\ .
\end{eqnarray}
Putting together conditions in Eqs.(\ref{eq:gauginobound}) and (\ref{eq:scalarbound}), 
we find that the $R$-invariant direct gauge mediation is successful only when
\begin{eqnarray}
\label{eq:scales}
\frac{k F}{M_{\rm mess}} &=& 10^{6-7}\,{\rm GeV}\ , \\
\frac{k F}{M_{\rm mess}^2} &\sim& 1\ .
\end{eqnarray}

In Fig.\,\ref{fig:spectrum}, we show a sample gauge mediated mass spectrum in the $R$-invariant direct gauge mediation model
for $N_M = 1$.
In the left plot, we show the spectrum as a function of $kF/(M_\Psi M_{\bar\Psi})$
while fixing $M_{\Psi} = M_{\bar \Psi} = 2\times 10^6$\,GeV and $kS_0/\sqrt{M_{\Psi}M_{\bar \Psi}} = 1$. 
Here we take $\tan\beta = 10$ although the SUSY spectrum barely depends on $\tan\b$.
In our analysis, we use {\tt SOFTSUSY 3.6.2}\,\cite{Allanach:2001kg} to calculate renormalization group
evolution of soft parameters as well as to analyze the electroweak symmetry breaking conditions.
The formulas of the gauge mediated spectrum at the messenger scale are given in Ref.~\cite{Nomura:1997uu,Sato:2009dk}.
As expected, the figure shows that gaugino masses become larger for a larger value of 
$kF/(M_\Psi M_{\bar\Psi})$, while the scalar masses are insensitive to this parameter.
It should be noted that the messenger scalars are tachyonic for $kF/(M_\Psi M_{\bar\Psi}) > 1$.
Therefore, the maximal gaugino masses are achieved for $kF/(M_\Psi M_{\bar\Psi}) \to 1$.

The right plot shows the mass spectrum as a function of $kS_0/\sqrt{M_{\Psi}M_{\bar \Psi}} $. 
Here we take $M_{\Psi} = M_{\bar \Psi} = 2\times 10^6$\,GeV and
fix $kF/(M_\Psi M_{\bar\Psi}) = 0.9$.
In the region of $kS_0/\sqrt{M_{\Psi}M_{\bar \Psi}} < 1$, gaugino masses increase 
with $kS_0/\sqrt{M_{\Psi}M_{\bar \Psi}}$,
while scalar masses are less sensitive to $kS_0/\sqrt{M_{\Psi}M_{\bar \Psi}}$.
This behavior comes from the fact that the gaugino masses require $R$-symmetry breaking,
while the scalar masses do not.
In the region of $kS_0/\sqrt{M_{\Psi}M_{\bar \Psi}} > 1$, both the gaugino masses and the 
scalar masses are decreasing with $kS_0/\sqrt{M_{\Psi}M_{\bar \Psi}}$.
This is because the messenger scale is dominated by $kS_0$ as in the conventional gauge mediation in that region.

For subsequent discussions, we split the messenger fields of $SU(5)$ GUT multiplets 
into the SM gauge group $SU(3)_C\times SU(2)_L\times U(1)_Y$ representations:
${\bf 5}=({\bf 3}_{\bf -1/3}, {\bf 2}_{\bf 1/2})$
and $\bar {\bf 5} = ({\bf \bar 3}_{\bf 1/3}, {\bf 2}_{\bf -1/2})$, 
such as $\Psi^{(\prime)} = (D^{(\prime)}, L^{(\prime)})$ and $\bar \Psi^{(\prime)} = (\bar D^{(\prime)}, \bar L^{(\prime)})$, respectively.
Accordingly, we also distinguish the parameters in Eq.\,(\ref{eq:SUSY-MESS}) 
for each messenger using subscripts: $k_D$ and $k_L$, $M_{D,\bar{D}}$ and $M_{L,\bar L}$, respectively.

In the analysis of Refs.~\cite{Sato:2009dk,Ibe:2012dd}, it is assumed that $k$'s and $M$'s satisfy the so-called GUT conditions at the GUT scale:
\begin{eqnarray}
\label{eq:GUT}
k_D = k_L\ , \quad M_D = M_L\ , \quad M_{\bar D} = M_{\bar L}\ .
\end{eqnarray}
In this paper, we do not impose these conditions in view of the 
fact that the doublets and the triplets of the Higgs multiplets in the GUT models are required to split.
In fact, the doublet-triplet splitting in the Higgs sector is most naturally achieved 
in GUT models with product gauge groups\,\cite{Izawa:1997he}.
In those models, the GUT conditions in Eq.\,(\ref{eq:GUT}) are not expected to be satisfied generically 
(see also Ref.~\cite{Harigaya:2015zea} for a recent discussion).
In the following, we simply take $k$'s and $M$'s of the $D$ and the $L$-type messengers as independent parameters.

It should be emphasized that the $R$-invariant direct mediation model is free from the $CP$-problem
from the messenger interactions.
The phases of $k_{D,L}$ and $M_{D,\bar{D}, L,\bar{L}}$ can be absorbed by
appropriate phase rotations of $D^{(\prime)}$, $\bar{D}^{(\prime)}$, $L^{(\prime)}$, and $\bar{L}^{(\prime)}$.
The phases of $S_0$ and $F$ can also be absorbed by the phases of $S$ and $\theta^2$, respectively.
In the above arguments, we have tacitly made use of these phase rotations to make $kF$'s, $kS_0$'s and $M$'s 
positive.%
\footnote{For $N_M \ge 2$, one may allow in Eq.\,(\ref{eq:SUSY-MESS}) couplings among fields with 
different labels $i$. 
Such couplings, however, lead to non-trivial phases on the parameters that result in 
relative phases to the gaugino masses and may bring about the SUSY $CP$-problem.
Those label-changing couplings can be suppressed by introducing a (approximate) $U(1)$ messenger symmetry for each label $i$, for example (see also Ref.~\cite{Ibe:2010ym}).
}

Before closing this subsection, let us comment on the upper limit on $N_M$ from the requirement 
of perturbative unification of gauge couplings.
In the $R$-invariant direct gauge mediation model, the $N_M = 1$ case includes 
two pairs of $({\bf 5},  \bar {\bf 5})$.
Besides, the messenger scale is at around $10^{6-7}$\,GeV as discussed above.
Therefore, the number of messengers in the messenger sector is severely constrained by the perturbative unification to $N_M \le 2$.
It should also be noted that the messenger fields in ${\bf 10}$ and ${\overline{\bf 10}}$ representations 
are also disfavored by the perturbative unification due to the doubled number of messengers
in the $R$-invariant direct gauge mediation.
In the following, we confine ourselves to the cases of $N_M = 1$ and $N_M = 2$ by taking the perturbative gauge coupling unification seriously.

\subsection{$B\mu$--Problem}

In the above analysis, we have not specified the origin of the $\mu$-term.
In fact, it is the long-sought problem about how to generate the $\mu$ and $B\mu$-terms of a similar size to other soft parameters while not causing the SUSY $CP$-problem.
The minimal possibility for the origin of the $\mu$-term is to assume that it is given just as is:
\begin{eqnarray}
\label{eq:mu}
W = \mu H_u H_d \ ,
\end{eqnarray}
where the $R$ charge of the two Higgs doublets is $2$.
As a notable feature of this type of $\mu$-term, the $B$-term at the messenger 
scale vanishes at the one-loop level:%
\footnote{A non-vanishing $B$-term is obtained from the two-loop threshold corrections 
of the messenger fields~\cite{Rattazzi:1996fb,Kahn:2013pfa} and is expected to be at around $10$\,GeV for the wino/bino masses around one TeV.}
\begin{eqnarray}
B \simeq 0 \ .
\end{eqnarray}
It should be also emphasized that this minimal model is favorable since it does not bring about the SUSY $CP$-problem.

One may consider more direct couplings between the Higgs doublets and the SUSY breaking sector to 
generate $\mu$ and $B$ terms, in order to interrelate the sizes of those parameters to other soft parameters.
Na{\"i}ve couplings between the SUSY breaking sector and the Higgs doubles, 
however, lead to too large a $B$-term, which is nothing but the infamous $\mu/B\mu$-problem.
More intricate connections between the Higgs and the SUSY breaking sector might be elaborated. 
In those models, one should be very careful to avoid the SUSY $CP$-problem.
 
In view of the minimality and the safety from the SUSY $CP$-problem, the minimal model of the $\mu$-term 
in Eq.\,(\ref{eq:mu}) seems to be the most favorable candidate.
In fact, many phenomenological studies have been done based on this minimal model of the $\mu$-term
in the conventional gauge mediation models~\cite{Rattazzi:1996fb,Gabrielli:1997jp,Hisano:2007ah}. 
As pointed out in Ref.~\cite{Ibe:2012dd}, however, the almost vanishing $B$-term at the messenger scale 
has a tension in the case of the $R$-invariant direct gauge mediation model as we see shortly.

In models with  the almost vanishing $B$-term at the messenger scale, 
the $B$ parameter at the stop mass scale is dominated by renormalization group effect:
\begin{eqnarray}
 \frac{d B}{d \ln \mu_R}  = 
 \frac{1}{16\pi^2} 
\left[
6 a_t y_t
+ 6 a_b y_b
+ 6 g_2^2 M_2 
+ \frac{6}{5}g_1^2 M_1
\right] \ ,
\end{eqnarray}
where $\m_R$ is the renormalization scale, $y_{t,b}$ the top and the bottom Yukawa coupling constants, and $a_{t,b}$ the corresponding trilinear soft parameters.  In the gauge mediation models, $a_{t,b}$ are also small and dominated by renormalization group effects from the gluino mass.
Roughly, the radiatively generated $B$-term at the stop mass scale is estimated to be
\begin{eqnarray}
\label{eq:RB}
| B(m_{\rm stop}) | \sim \frac{1}{16\pi^2} 
\left[
6 a_t y_t
+ 6 a_b y_b+
 6 g_2^2 M_2 
+ \frac{6}{5}g_1^2 M_1
\right]  \log\frac{M_{\rm mess}}{m_{\rm stop}}
< O(0.1) \times M_2 \ ,
\end{eqnarray}
where we have taken the messenger scale, $M_{\rm mess } = O(10^{6-7})$\,GeV, 
and $m_{\rm stop} \sim 10$\,TeV.
In our analysis, we use the convention that gaugino masses, $B\mu$, and $\tan\beta$ are positive-valued.
From Eq.\,(\ref{eq:RB}), the radiatively generated $B(m_{\rm stop})$ is negative-valued
at the low energy scale.
Thus,  the sign of $\mu$ is negative in our convention.

The radiatively generated $B$-term is, generically, too small and renders too large a $\tan\b$:
\begin{eqnarray}
\label{eq:tanb}
\tan\b  \simeq \frac{2}{\sin2\b}
\simeq \frac{m_{H_u}^2 + m_{H_d}^2 + 2 \m^2} {B(m_{\rm stop}) \m} 
\simeq \frac{m_{H_d}^2 + |m_{H_u}^2|} {|B(m_{\rm stop})| |m_{H_u}|}  
> \frac{2 m_{H_d}}{|B(m_{\rm stop})|} > {\cal O}(100)\ .
\end{eqnarray}
Such a large $\tan\b$ leads to too large a bottom Yukawa coupling.%
\footnote{More precisely, the Standard Model down-type Yukawa couplings are 
dominated by the radiative generated non-holomphic coupling to $H_u$\,\cite{Dobrescu:2010mk}
for such a large $\tan\b$ (see also\,\cite{Altmannshofer:2010zt,Bach:2015doa}).
In this paper, we confine ourselves to the case where the down-type Yukawa couplings
come from the coupling to $H_d$ in the superpotential.
}
In the third equality of Eq.\,(\ref{eq:tanb}), we have used the electroweak symmetry breaking for a large $\tan\b$,
\begin{eqnarray}
\mu^2 = \frac{-m_{H_u}^2 \tan^2\b+ m_{H_d}^2}{\tan^2\beta - 1} + \frac{1}{2}m_Z^2 \simeq |m_{H_u}^2|\ .
\label{eq:musq}
\end{eqnarray}
We have also used $m_{H_u}^2 < 0 $ and $m_{H_d}^2 > |m_{H_u}^2|$ which is valid for most parameter space.
In the final inequality, we have used Eqs.\,(\ref{eq:gaugino}) and (\ref{eq:RB}).
It should be emphasized that this tension is due to the hierarchy between the gaugino mass
and the scalar mass in the $R$-invariant direct gauge mediation.%
\footnote{The expected size of $\tan\b$ is smaller for $N_M = 2$ compared with the case for $N_M = 1$,
as the relative size of the gaugino mass (especially $M_2$) to the scalar mass becomes larger 
for $N_M = 2$.
}

\begin{figure}[t]
\begin{center}
\begin{minipage}{.46\linewidth}
  \includegraphics[width=\linewidth]{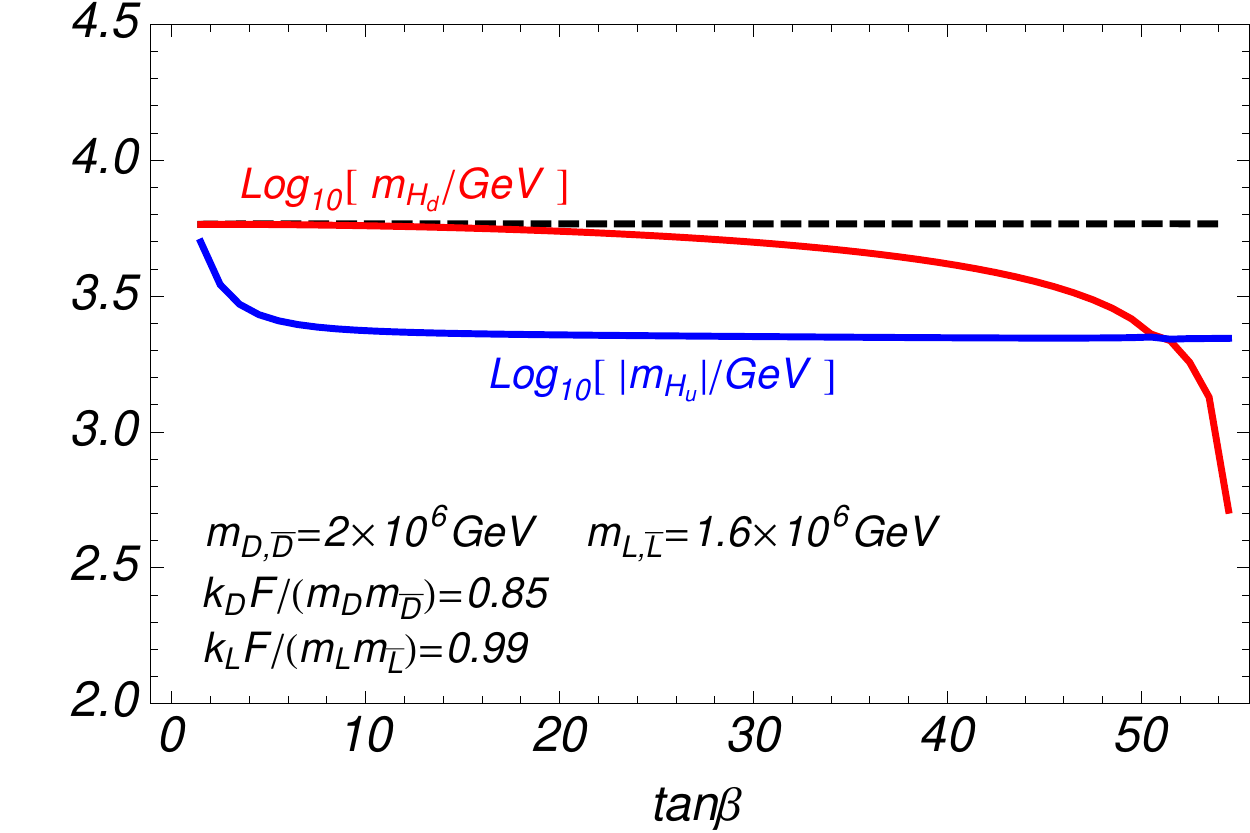}
 \end{minipage}
 \hspace{1cm}
 \begin{minipage}{.46\linewidth}
  \includegraphics[width=\linewidth]{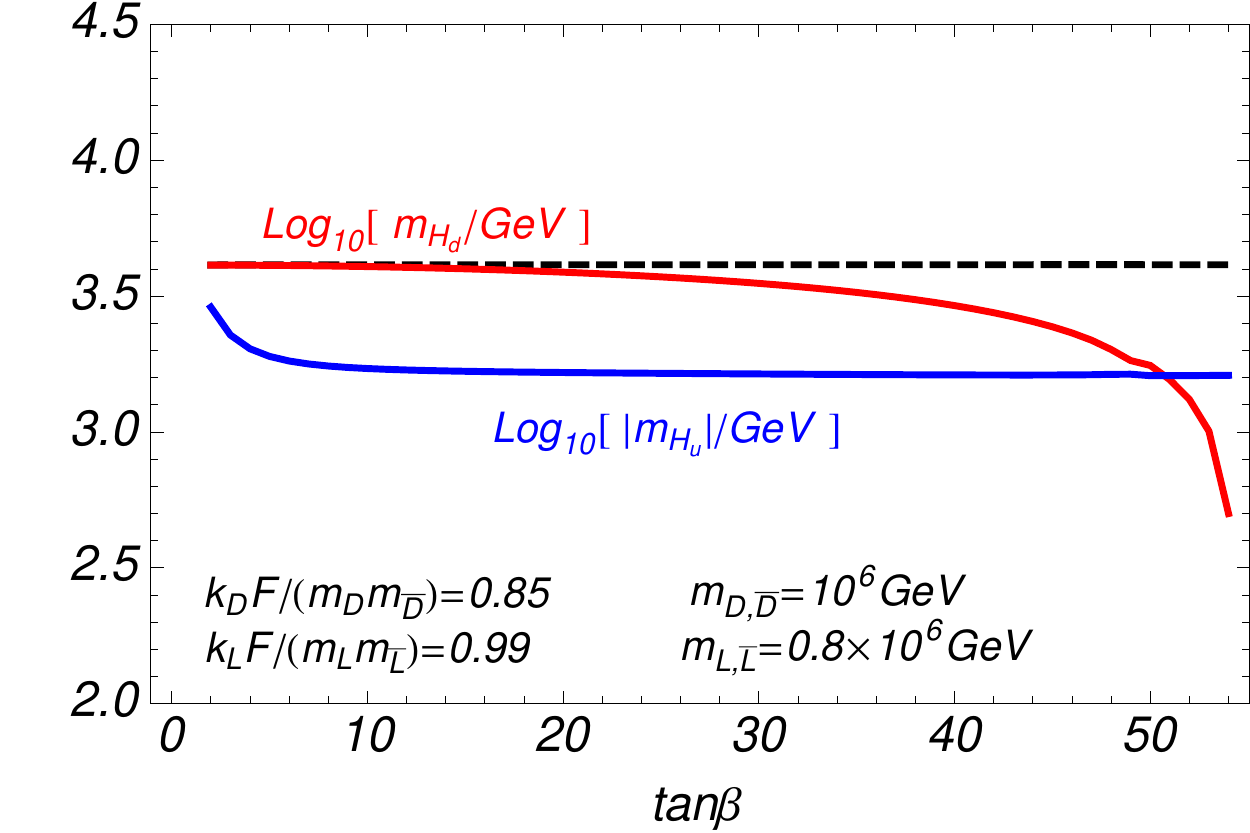}
 \end{minipage}
 \end{center}
\caption{\sl \small
The soft masses of $H_u$ and $H_d$ as  functions of $\tan \beta$
for $N_M = 1$ (left) and $N_M = 2$ (right).
The red and blue curves are $m_{H_d}$ and $|m_{H_u}|$ ($m_{H_u}^2 <0$) 
at the stop mass scale, respectively. 
The dashed curves are the corresponding soft masses at the messenger scale.
The other parameters are explicitly indicated in each plot.
}
\label{fig:softMH}
\end{figure}

The above generic argument has a loophole.
That is, we have assumed $m_{H_d}^2(m_{\rm stop}) \simeq m_{H_d}^2(M_{\rm mess})$.
Although this relation is valid in most parameter space, 
it becomes invalid when 
$\tan\b \gtrsim 50$ and the bottom Yukawa coupling $y_b$ 
becomes comparable to the top Yukawa coupling.
For such a large $y_b$, $m_{H_d}^2$ also receives a sizable
negative contribution from the sbottom soft masses and gets
smaller at the lower energy scale as $m_{H_u}^2$ does.
When $m_{H_d}^2(m_{\rm stop})\ll m_{H_d}^2(M_{\rm mess})$ is achieved, 
the resultant $\tan\beta$ can be much smaller than the one expected in Eq.\,(\ref{eq:tanb})
and within a viable range.
In this way, the $R$-invariant direct gauge mediation model can become consistent with
the boundary condition with $B(M_{\rm mess}) \simeq 0$.

In Fig.\,\ref{fig:softMH}, we show  $m_{H_d}$ and $|m_{H_u}|$ as functions of $\tan \b$.
The plots show that  $m_{H_d}^2(m_{\rm stop}) \simeq m_{H_d}^2(M_{\rm mess})$ for a moderate value of $\tan\b$ as expected.
For  $\tan \b \gtrsim 50$, on the other hand, the renormalization group effects on $m_{H_d}^2$ 
are sizable, and its becomes much smaller at the stop mass scale than at the messenger scale.

\begin{figure}[t]
\begin{center}
\begin{minipage}{.46\linewidth}
  \includegraphics[width=\linewidth]{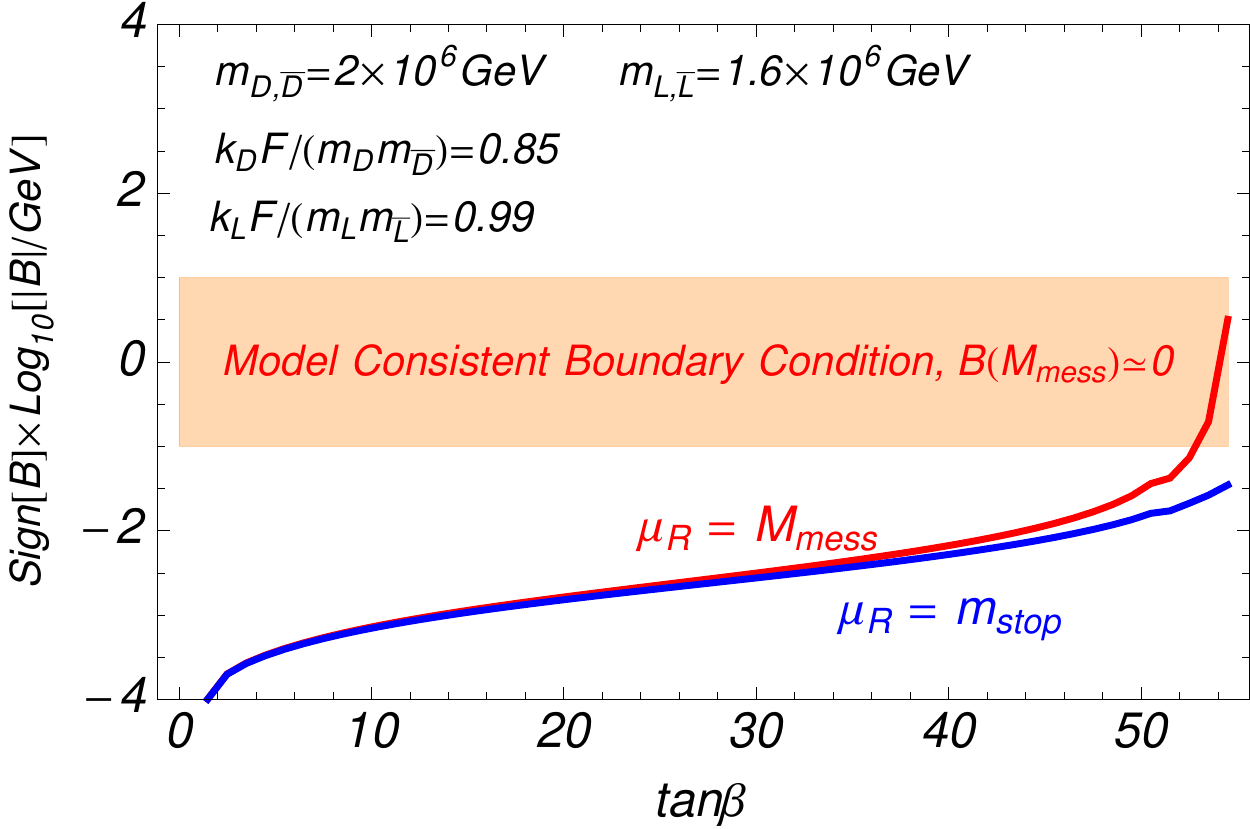}
 \end{minipage}
 \hspace{1cm}
 \begin{minipage}{.46\linewidth}
  \includegraphics[width=\linewidth]{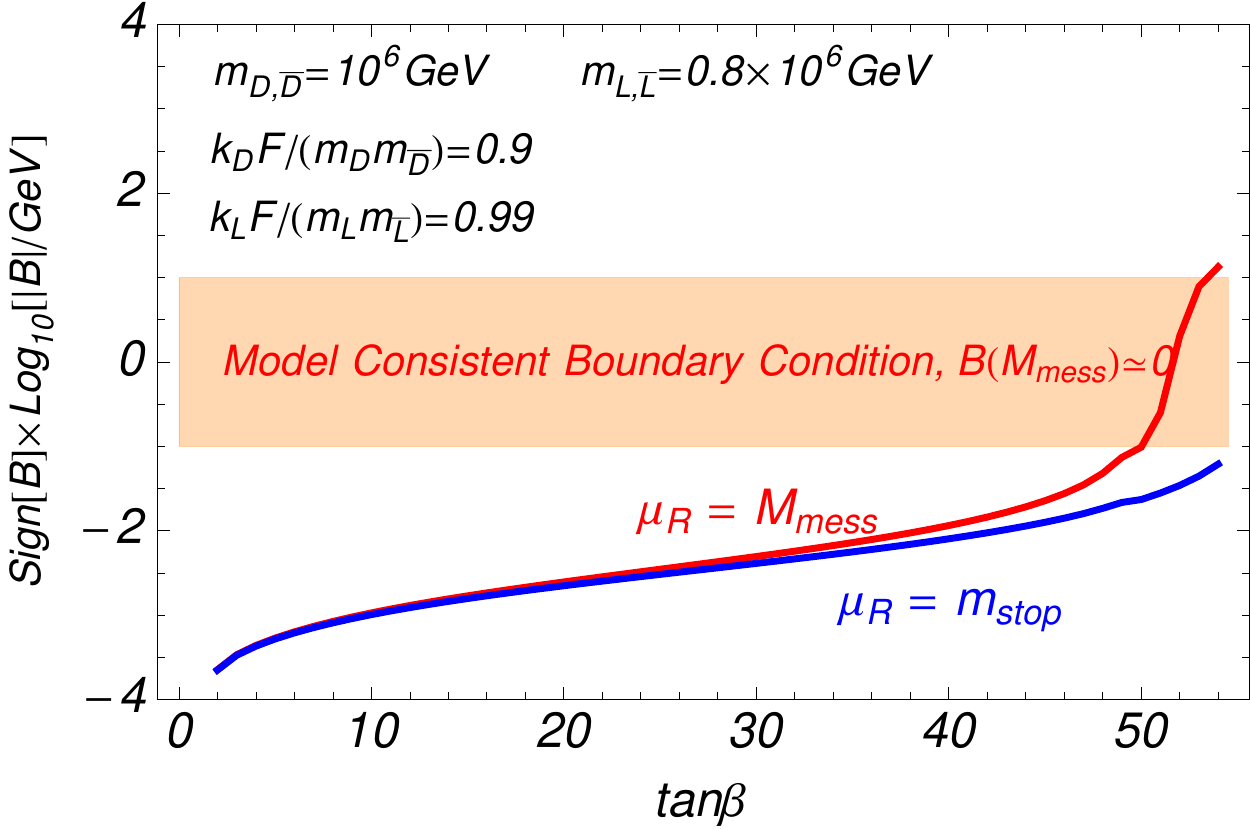}
 \end{minipage}
 \end{center}
\caption{\sl \small
The $B$-term at the messenger scale (red curves) and the stop mass scale (blue curves) as functions of $\tan\beta$
for $N_M = 1$ (left) and $N_M = 2$ (right).
The other parameters are indicated explicitly in each plot.
The boundary condition $B(M)\simeq 0$ is satisfied for $\tan\b\gtrsim 50$.
}
\label{fig:B}
\end{figure}

Armed with this observation, we have searched the parameter space for regions where the $R$-invariant 
direct gauge mediation is consistent with the boundary condition with an almost vanishing $B$-term.
Fig.\,\ref{fig:B} shows the $B$-term at the messenger scale (red curves) 
and the stop mass scale (blue curves) as functions of $\tan\beta$ in the $R$-invariant direct gauge mediation model.
Here we impose the electroweak symmetry breaking condition with $m_Z \simeq 91.2$\,GeV, instead of the boundary condition  $B(M_{\rm mess})\simeq 0$.
The plots show that the boundary condition, $B(M_{\rm mess})\simeq 0$, 
is compatible with the $R$-invariant model for $\tan \beta \gtrsim 50$, as expected.
As a result, we find that the $R$-invariant direct gauge mediation model is consistent with the minimal $\mu$-term.

It should be emphasized again that the consistency between the $R$-invariant direct gauge mediation
model and the boundary condition $B(M_{\rm mess})\simeq 0$ is more difficult  than 
in the case of the conventional gauge mediation.
This difficulty stems from the hierarchy between the gaugino mass and the scalar mass
as well as from the low messenger scale, $M \simeq 10^{6-7}$\,GeV (see Eq.\,(\ref{eq:scales})).
In usual gauge mediation models, the messenger scale can be much larger while keeping the 
soft breaking mass scales in the TeV range, with which the radiatively generated $B$-term can be sizable
due to a rather long interval of the renormalization group running.
In the $R$-invariant direct gauge mediation model, on the other hand,  
one needs to take $k_L F/(M_L M_{\bar{L}})$ to be very close to $1$, so that 
the gaugino mass, $M_2$, takes a value as large as possible with which the 
the radiatively generated $B$-term at the low energy becomes sizable.

\begin{figure}[t]
\begin{center}
\begin{minipage}{.46\linewidth}
  \includegraphics[width=\linewidth]{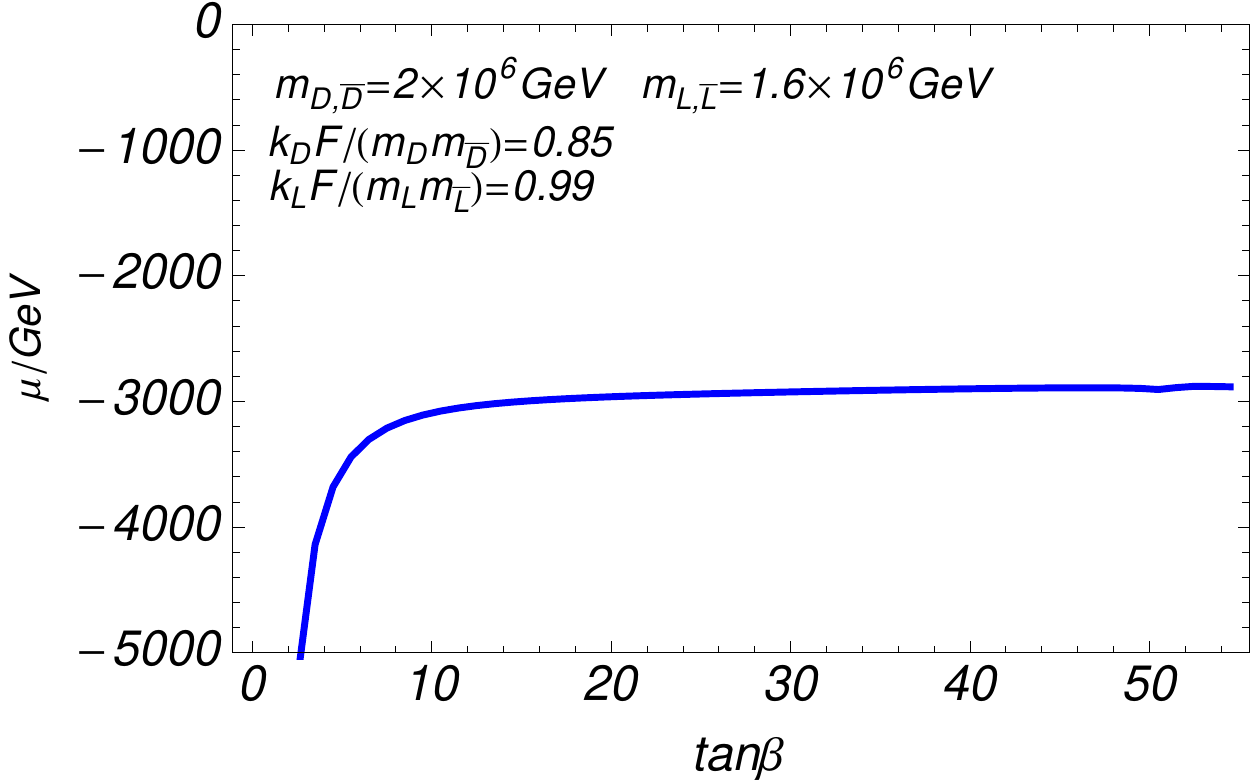}
 \end{minipage}
 \hspace{1cm}
 \begin{minipage}{.46\linewidth}
  \includegraphics[width=\linewidth]{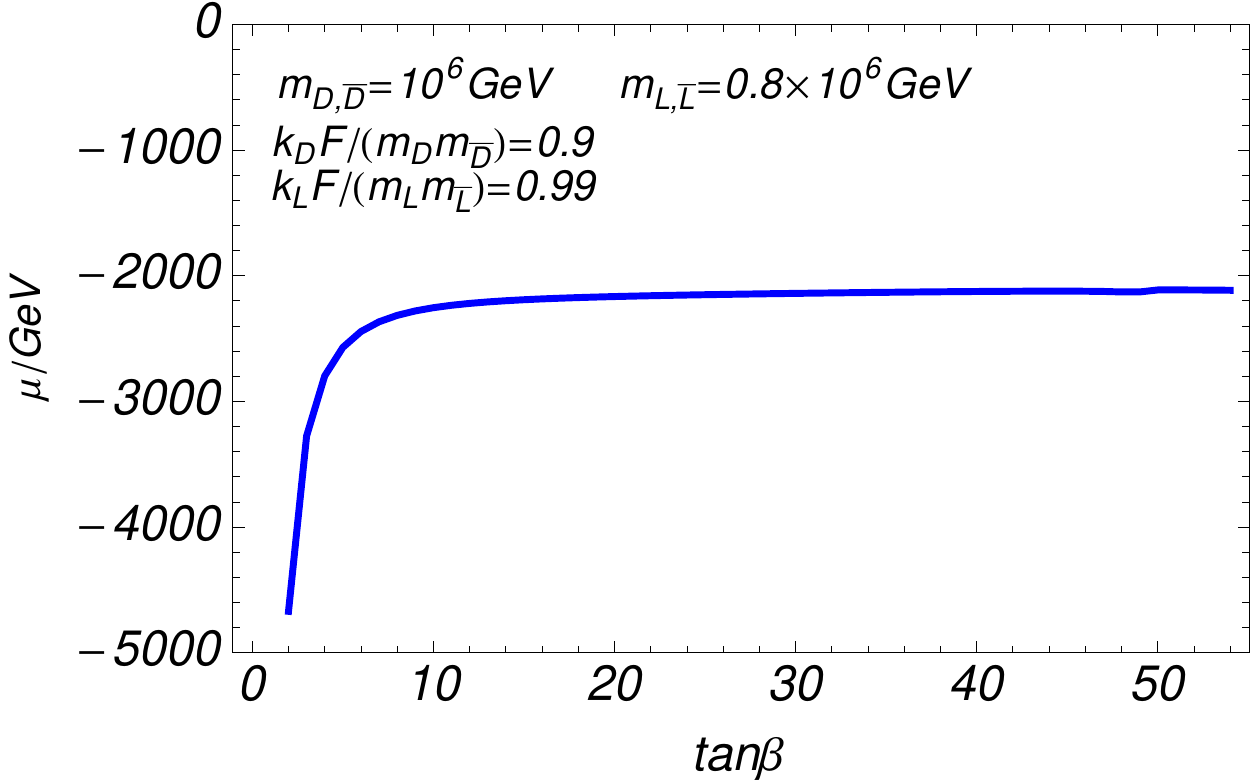}
 \end{minipage}
 \end{center}
\caption{\sl \small
The required $\mu$-term as functions of $\tan\beta$ for $N_M = 1$ (left) and $N_M = 2$ (right).
The other parameters are indicated explicitly in each plot.
The corresponding stop mass scales are $m_{\rm stop} \simeq 12$\,TeV ($N_M=1$)
and $m_{\rm stop} \simeq 9$\,TeV ($N_M=2$), respectively.}
\label{fig:mu}
\end{figure}

Before closing this subsection, let us comment on the required size of $\mu$
for successful electroweak symmetry breaking.
As we have argued in Eq.\,(\ref{eq:musq}), the required size of $\mu$ is roughly given by
\begin{eqnarray}
\mu^2 \simeq - m_{H_u}^2(m_{\rm stop}) 
\end{eqnarray}
for a large $\tan \b$.
Here the Higgs soft mass squared, $m_{H_u}^2$, is approximately given by
\begin{eqnarray}
m_{H_u}^2(m_{\rm stop})  \sim m_{H_u}^2(M_{\rm mess}) - \frac{12  y_t^2} {16\pi^2} \,m_{\rm stop}^2 \,\log\frac{M_{\rm mess}}{m_{\rm stop}}
\end{eqnarray}
at the stop mass scale, which can be much smaller than $m_{\rm stop}\simeq 10$\,TeV
for $M_{\rm mess} \simeq 10^{6-7}$\,GeV.
As a result, the required size of $\mu$-term is also much smaller than $m_{\rm stop}$.
This feature somewhat eases the electroweak fine-tuning problem 
while explaining the observed Higgs boson mass by a heavy stop mass of ${\cal O}(10)$\,TeV.
In Fig.\,\ref{fig:mu}, we show the required size of the $\mu$-term as a function of $\tan\beta$ by taking 
the same parameter sets used in Fig.\,\ref{fig:B}.
The figure shows that the required $\mu$-term is indeed smaller than the stop mass scale.
It should be also noted that a smaller $\mu$-term is also possible when $m_{H_u}^2(M_{\rm mess})$ 
is slightly larger at the messenger scale.
This property is important for the discussions in section~\ref{sec:signal}.

\subsection{Gravitino Dark Matter}

In  gauge mediation models, the gravitino is the lightest supersymmetric particle (LSP). 
By assuming $R$-parity conservation, it can serve as a  candidate for dark matter.
In the regime of much lighter than MeV, the gravitino is thermalized 
in the early Universe, and its relic abundance is estimated to be
\begin{eqnarray}
\Omega_{3/2} h^2 \simeq 0.1 \left(\frac{100}{g_*(T_D)}\right) \left(\frac{m_{3/2}}{100~{\rm eV}}\right) .
\label{eq:abundance}
\end{eqnarray}
Here $m_{3/2}$ is the gravitino mass, and $g_*(T_D) \simeq 100$ denotes the effective massless degree of freedom in the thermal bath at the decoupling temperature\,\cite{Moroi:1993mb}
\begin{eqnarray}
T_{D}\sim \max\left[
M_{\tilde g},
160\,{\rm GeV}\left(\frac{g_*(T_D)}{100}\right)^{1/2}
\left(\frac{m_{3/2}}{10\,{\rm keV}}\right)^2
\left(\frac{2\,{\rm TeV}}{M_{3}}\right)^2
\right]\ .
\end{eqnarray}

As discussed above, a successful $R$-invariant direct gauge mediation requires
\begin{eqnarray}
(k F)^{1/2} = 10^{6-7}\, {\rm GeV}\ .
\end{eqnarray}
By assuming that the SUSY breaking field $S$ breaks supersymmetry dominantly, 
the gravitino mass is given by
\begin{eqnarray}
m_{3/2} \simeq 10\,{\rm keV}\times \left(\frac{0.1}{k}\right)\left( \frac{(k F)^{1/2}}{2\times10^6\,{\rm GeV}}\right)^2\ .
\end{eqnarray}
In this case, the thermally produced gravitino abundance in Eq.\,(\ref{eq:abundance}) is too large to be consistent 
with the observed dark matter density.%
\footnote{The gravitino with a mass $m_{3/2}\simeq 100$\,eV is 
not cold dark matter but hot dark matter.  Hence, it is not a viable candidate for dark matter
even if the thermal relic abundance is consistent with the observed dark matter density.}

This tension is removed when the above relic density is diluted by entropy production 
by a factor of
\begin{eqnarray}
\Delta \simeq 100\times \left(\frac{100}{g_*(T_D)}\right)
\left(\frac{m_{3/2}}{10\,\rm keV}\right)
\end{eqnarray}
after the gravitino decouples from the thermal bath.%
\footnote{In general, if the dilution factor is provided by a late-time decay of a massive 
particle which dominates the energy density of the Universe, it is given by
$\Delta \simeq T_{\rm dom}/T_{\rm decay}$, where $T_{\rm dom}$ is the temperature at which 
the massive particle dominates the energy density of the Universe and $T_{\rm decay}$ is its decay temperature.
In order not to affect the Big-Bang Nucleosynthesis, we require $T_{\rm decay} \gtrsim {\cal O}(1$--$10)$\,MeV,
and hence the dilution factor is bounded from above by $\Delta < T_{\rm dom}/{\cal O}(1\mbox{--}10)\,\rm MeV$.}
As shown in Ref.~\cite{Ibe:2010ym}, an appropriate amount of entropy can be provided by, for example, the decay of long-lived particles in the dynamical SUSY breaking sector.%
\footnote{A mass of $10^{6-7}$\,GeV for the messenger is too light to provide a sufficient dilution factor\,\cite{Fujii:2002fv} (see also appendix \ref{sec:Rcharge}).
For other mechanisms of entropy production after the decoupling of gravitinos,
see Refs\,\cite{Fujii:2003iw, Hasenkamp:2010if}. }
Interestingly, the gravitino in this mass range is a good candidate for a slightly warm dark matter\,\cite{Ibe:2012dd} enabled via an appropriate dilution factor.

Finally, let us  comment on the decay length of the NLSP, which is the bino in most parameter space of the $R$-invariant direct gauge mediation model.
The bino NLSP mainly decays into a gravitino and a photon/$Z$-boson with the branching ratios
\begin{eqnarray}
\G(\tilde b \to  \psi_{3/2} + \g) &\simeq& \frac{1}{48\pi}
\frac{M_{1}^5}{M_{PL}^2 m_{3/2}^2} \cos^2\theta_W\ , \\
\G(\tilde b \to  \psi_{3/2} + Z) &\simeq& \frac{1}{48\pi}
\frac{M_{1}^5}{M_{PL}^2 m_{3/2}^2} \sin^2\theta_W\ . 
\end{eqnarray}
Here $M_{\rm PL} \simeq 2.4 \times 10^{18}$\,GeV denotes the reduced Planck scale,
and $\theta_W$ is the weak mixing angle.
Altogether, the decay length of the bino NLSP is given by
\begin{eqnarray}
\label{eq:length}
c\tau_{\tilde B} \simeq 0.6\,{\rm m}\times \left(\frac{500\,{\rm GeV}}{M_{1}}\right)^5\left(\frac{m_{3/2}}{10\,\rm{keV}}\right)^2\ .
\end{eqnarray}
Therefore, the bino may or may not decay inside the detectors, depending on the gravitino mass and the NLSP mass.

\section{Upper bound on the gluino mass}\label{sec:gluino}

As alluded to before, the $R$-invariant direct gauge mediation model predicts a hierarchy between gaugino masses 
and scalar masses.  We have also argued that $\tan\beta$ is required to be large, $\tan\beta \gtrsim 50$,
if we further assume that the $\mu$-term is provided by the minimal $\mu$-term in the superpotential, Eq.\,(\ref{eq:mu}).
For such a large $\tan\beta$, the stop mass is restricted to be around 10\,TeV to account for the observed Higgs boson mass,
$m_h = 125.09\pm 0.21\pm 0.11$\,GeV\,\cite{Aad:2015hea}. 
Thus, by remembering that the gluino mass is limited from above for a given squark mass, 
the observed Higgs boson mass  leads to an upper bound on the gluino mass.

\begin{figure}[t]
\begin{center}
  \includegraphics[width=.6\linewidth]{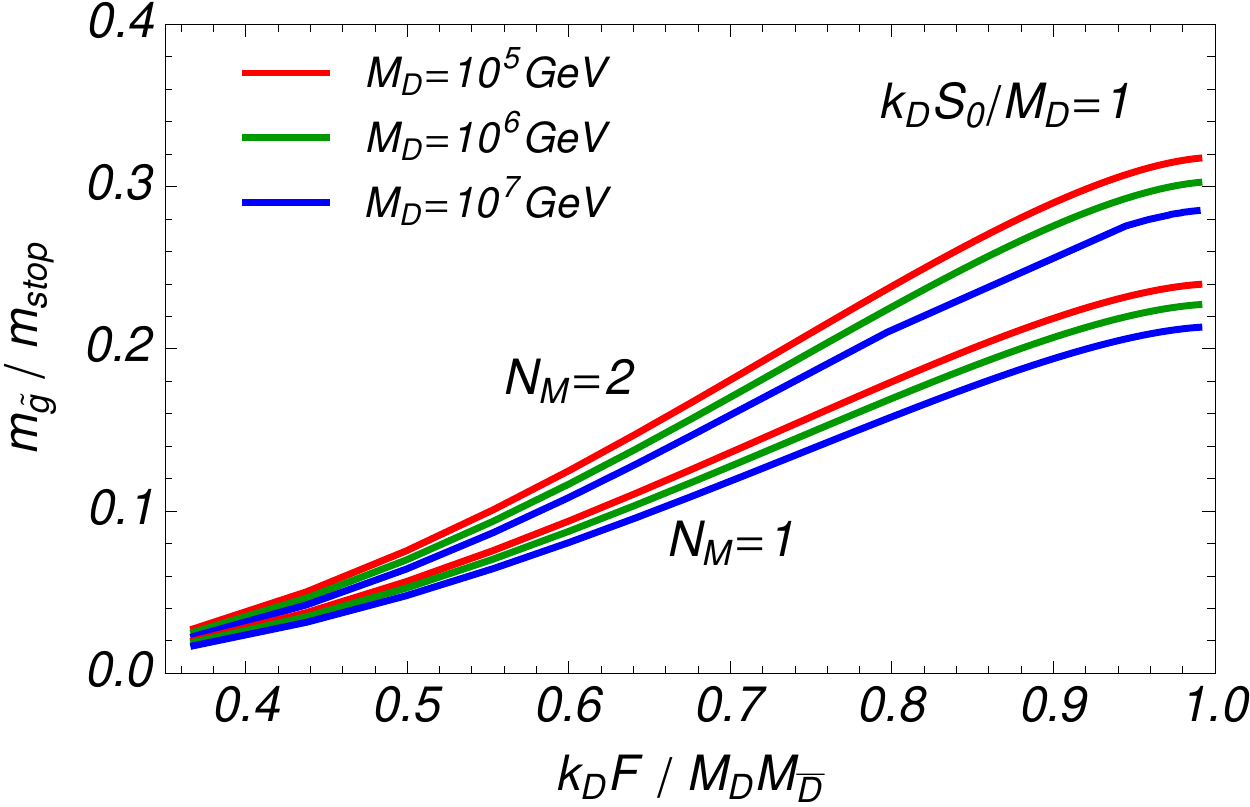}
 \end{center}
\caption{\sl \small
The ratio between the gluino mass and the stop mass, $m_{\rm stop}\equiv \sqrt{m_{\tilde{t}_1}m_{\tilde{t}_2}}$,
as a function of $k_D F/(M_D M_{\bar D})$ for $N_M = 1$ and $N_M = 2$.
In the figure, we fix $M_D = M_{\bar D}$ and $k_D S_0 = M_D$
in order to optimize the gluino mass.}
\label{fig:ratio}
\end{figure}

To obtain the limit on the gluino mass from the observed Higgs boson mass, 
let us first consider the ratio between the gluino mass and the stop mass, which is shown as a function $k_D F/(M_D M_{\bar D})$ for $N_M= 1$ and $N_M = 2$ in Fig.\,\ref{fig:ratio}.
Here we take $k_D S_0 = (M_D M_{\bar D})^{1/2}$ to maximize the gluino mass, 
as given in Fig.\,\ref{fig:spectrum}.
We have also checked that a heavier gluino mass cannot be achieved even for $M_D \neq M_{\bar D}$.
The figure indicates that the ratio is a monotonically increasing function of $k_D F/(M_D M_{\bar D})$ and that $m_{\tilde g}/m_{\rm stop}\lesssim 0.2$ for $N_M =1$ and $m_{\tilde g}/m_{\rm stop}\lesssim 0.3$ for $N_M =2$.

\begin{figure}[t]
\begin{center}
\begin{minipage}{.46\linewidth}
  \includegraphics[width=\linewidth]{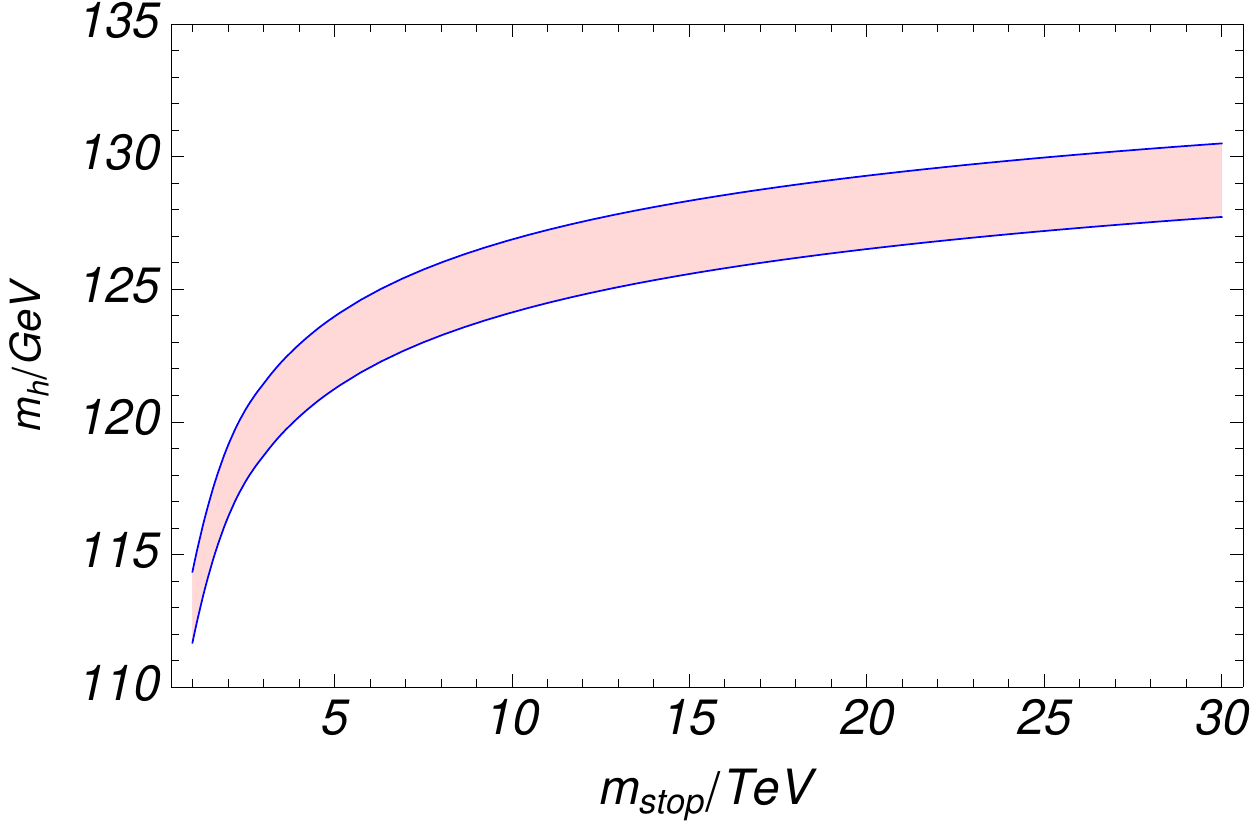}
 \end{minipage}
 \hspace{1cm}
 \begin{minipage}{.46\linewidth}
  \includegraphics[width=\linewidth]{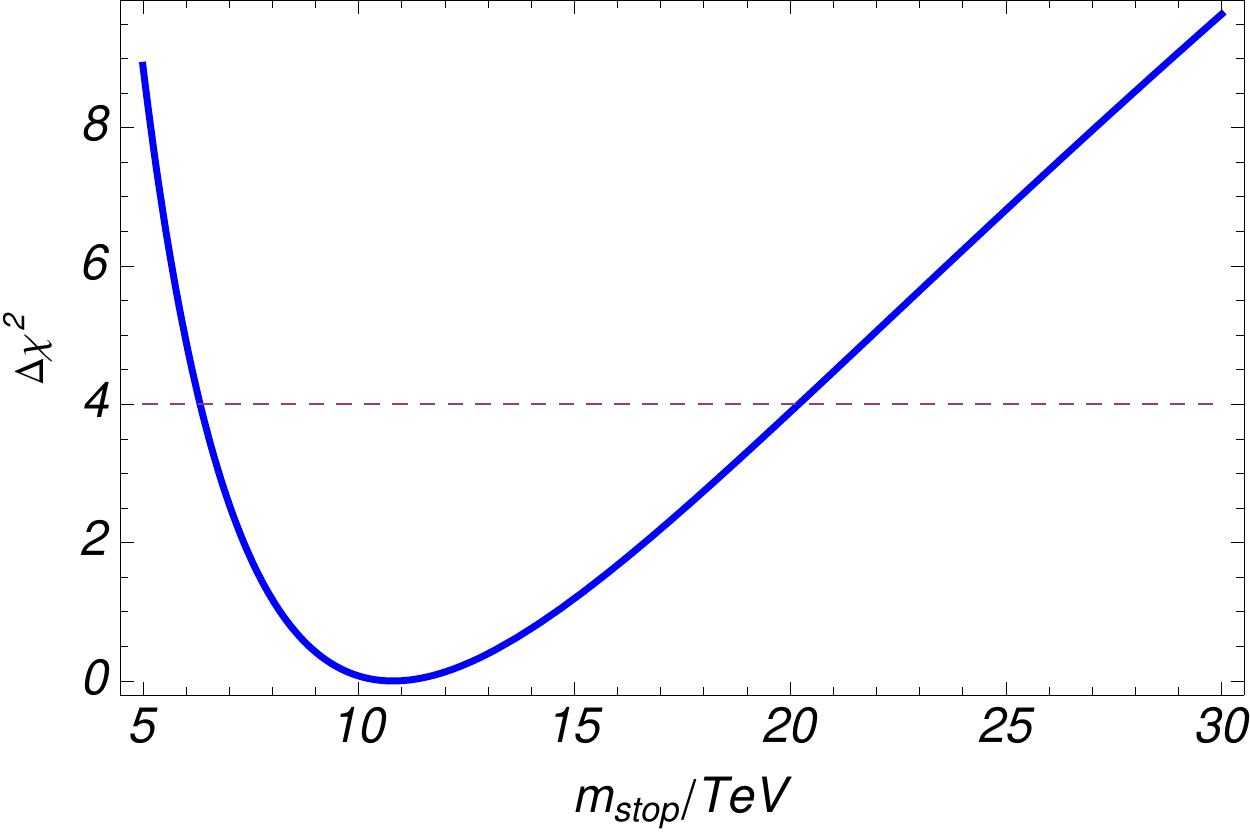}
 \end{minipage}
 \end{center}
\caption{\sl \small
The predicted Higgs boson mass as a function of 
the stop mass (left).
${\mit \D}\chi^2$ as a function of the stop mass (right).
The band in the left plot shows the theory uncertainties.
}
\label{fig:Higgs}
\end{figure}

From the observed Higgs boson mass, the stop mass is restricted to be in the ${\cal O}(10)$~TeV regime.
The left plot of Fig.\,\ref{fig:Higgs} shows the Higgs boson mass as a function of the stop mass for $\tan\b \gtrsim 50$.
In our analysis, we use {\tt SusyHD}~\cite{Vega:2015fna} to calculate the Higgs boson mass.
The band represents uncertainties of our prediction, originating from (i) higher-order corrections of the Higgs mass calculation,
(ii) uncertainties of Standard Model input parameters, and (iii) choices of model parameters that affect the SUSY
 spectrum other than stop masses.
In our analysis, we take the first uncertainty to be $1$\,GeV to make our discussions conservative (see also Ref.~\cite{Vega:2015fna}). 
As for the uncertainties from Standard Model parameters, the top mass uncertainty $m_t = 173.21 \pm 0.51\pm0.71$\,GeV
is the most important one for the Higgs mass prediction, and amounts to an error of about $0.3$\%.
The effect of model parameters to which the stop mass is negligible. 
In the right plot, we show the stop mass dependence of ${\mit \D}\chi^2$ defined by
\begin{eqnarray}
{\mit \D} \chi^2 = \frac{(125.09-m_h)^2}{\sqrt{\s_{\rm ex}^2 + \s_{\rm th}^2}}\ ,
\end{eqnarray}
where $\s_{\rm ex}$ denotes the experimental error and $\s_{\rm th}$ denotes the theoretical error as listed above.
The $1\,\s$ ($2\,\s$) upper limit corresponds to $m_h \simeq 126.2$\,GeV ($127.2$\,GeV).
From the plot, we find that $m_{\rm stop} \gtrsim 21$\,TeV is excluded at $2\sigma$ level (as indicated by the horizontal dashed line).%
\footnote{If we use {\tt FeynHiggs 2.11.2} \cite{Heinemeyer:1998yj}, we obtain a Higgs boson mass larger by about $4$\,GeV for $m_{\rm stop}$ in the ${\cal O}(10)$ TeV regime.  This leads to a more stringent bound on the gluino mass.}
Thus, by remembering that $m_{\tilde g}/m_{\tilde t} \lesssim 0.2$ for $N_M = 1$  and 
$m_{\tilde g}/m_{\tilde t} \lesssim 0.3$ for $N_M = 2$, 
we immediately find respectively $m_{\tilde g} \lesssim 4$\,TeV and $m_{\tilde g} \lesssim 6$\,TeV.

\begin{figure}[t]
\begin{center}
\begin{minipage}{.46\linewidth}
  \includegraphics[width=\linewidth]{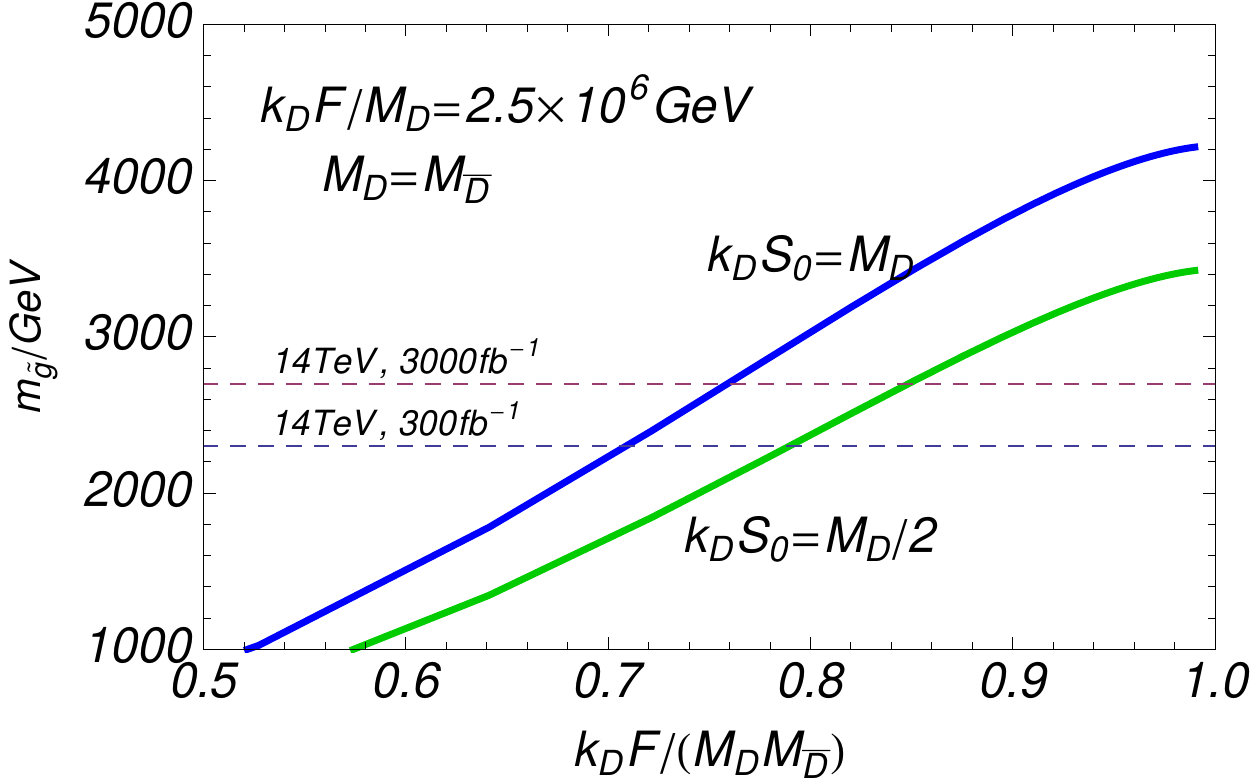}
 \end{minipage}
 \hspace{1cm}
 \begin{minipage}{.46\linewidth}
  \includegraphics[width=\linewidth]{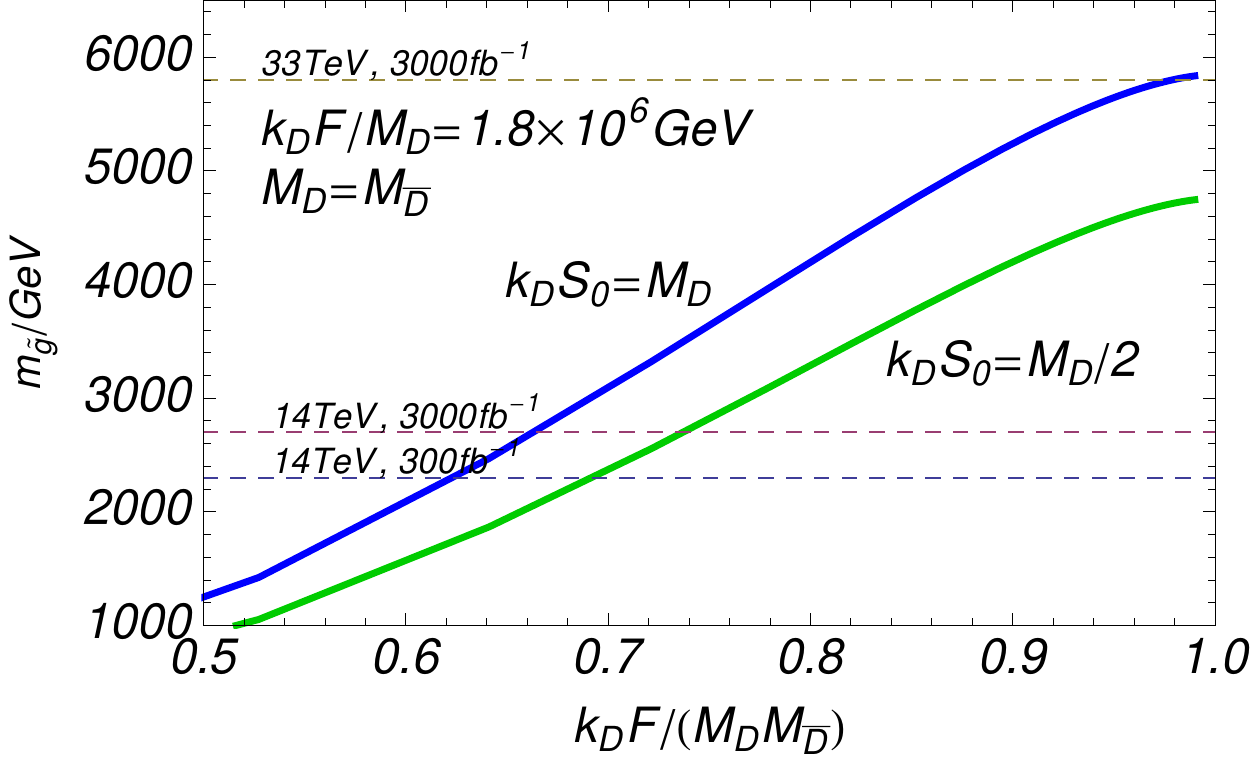}
 \end{minipage}
 \end{center}
\caption{\sl \small
The upper limit on the gluino mass as a function of $k_D F/(M_D M_{\bar D})$
for $N_M = 1$ (left) and $N_M = 2$ (right).
Each plot corresponds to the stop mass around 20\,TeV, the upper limit obtained 
from the Higgs boson mass.
The blue curves show the upper limits for an optimal $R$-symmetry breaking, $k_D S_0 = (M_DM_{\bar D})^{1/2}$,
and the green curves show those for a less optimal $R$-symmetry breaking, $k_D S_0 = (M_DM_{\bar D})^{1/2}/2$.
The dashed lines are the expected 95\%\,CL limits on the gluino mass\,\cite{Cohen:2013xda}.
}
\label{fig:gluino}
\end{figure}

For a closer look, we show in Fig.\,\ref{fig:gluino} the gluino mass as a function of $k_D F/(M_D M_{\bar D})$
for $k_DF/(M_D M_{\bar D})^{1/2} \simeq 2.5\times 10^6\,$GeV and 
$N_M = 1$ and for $k_DF/(M_D M_{\bar D})^{1/2} \simeq 1.8\times 10^6\,$GeV and $N_M = 2$.
Such parameter choices correspond to the upper limit of the stop mass, $m_{\rm stop} \simeq 20$\,TeV 
for each $N_M$.
In the figure, we also show the expected 95\%\,CL lower limits on the gluino mass at the 14-TeV LHC with the integrated luminosity $300$\,fb$^{-1}$ and $m_{\tilde g} < 2.3$\,GeV,
at the HL-LHC with the integrated luminosity $3000$\,fb$^{-1}$ and $m_{\tilde g} < 2.7$\,GeV,
and at a 33-TeV hadron collider with the integrated luminosity $3000$\,fb$^{-1}$ and $m_{\tilde g} < 5.8$\,GeV~\cite{Cohen:2013xda}.
The figure shows that the LHC experiment will cover a large portion of the parameter space
for $N_M = 1$ unless the gluino mass is highly optimized.
The figure also shows that a 33-TeV hadron collider will cover almost the entire gluino mass range.

\section{$Z$+jets+missing $E_T$}\label{sec:signal}

Recently, the ATLAS Collaboration reported a $3\sigma$ excess in the search
for events with a $Z$ boson (decaying into a lepton pair) accompanied by jets and a large missing transverse energy
($E_T^{\rm miss}$)\,\cite{Aad:2015wqa}. 
They observed 29 events in a combined signal region with di-electrons and di-muons at the $Z$-pole
in comparison with an expected background of $10.6\pm 3.2$ events.
Although the significance of the signal is not sufficiently high at this point, many attempts have been made 
to explain the excess using the MSSM\,\cite{Lu:2015wwa,Barenboim:2015afa,Kobakhidze:2015dra,Cahill-Rowley:2015cha,Collins:2015boa}. 
The signal requires colored SUSY particles lighter than about 1.2\,TeV\,\cite{Barenboim:2015afa}.
In this section, we briefly discuss whether the $R$-invariant direct gauge mediation model can explain the reported signal.

As discussed in the previous section, the $R$-invariant direct gauge mediation model predicts a mass hierarchy between
the gauginos and the scalars.  Hence, the candidate colored SUSY particle required 
for the signal is inevitably the gluino.
For such a light gluino, constraints from searches for SUSY particles with jets$+E_T^{\rm miss}$
are usually severe, and most parameter region has been excluded~\cite{Khachatryan:2015vra,Aad:2015iea}. 
As shown in Ref.~\cite{Lu:2015wwa}, however, a careful study shows that the reported signal can be 
explained by the gluino production while evading all the other constraints.
In this case, the gluino decays mainly into a gluon and a neutral Higgsino which subsequently decays
into a $Z$-boson and a stable bino.
Such a two-body decay of the gluino is induced by top-stop-loop diagrams,
which can be dominant when the mass difference between the gluino and the Higgsino is sufficiently small 
and squark masses are much heavier than gaugino masses\,\cite{Lu:2015wwa}. 
The dominance of the two-body decay mode is important to evade constraints from SUSY searches of multi-jets$+E_{T}^{\rm miss}$.
It is also important to suppress the decay of the neutral Higgsino into a Higgs boson and a bino
by requiring $M_{\tilde H} - M_{\tilde b} \lesssim 100$\,GeV.
In addition, preventing the wino from appearing in the gluino decay chain is also important to avoid constraints from other SUSY searches.

\begin{table}
\begin{center}
\begin{tabular}{|| l | cl || l | cl ||} 
\hline
 ${\tilde g}$       & 810&GeV      & ${\tilde{u}_L}$  & 14&TeV   \\
 ${\chi_1^0(\tilde b) }$      & 521&GeV      & ${\tilde{u}_R}$  & 13&TeV   \\
 ${\chi_2^0(\tilde H)}$        & 623&GeV     & ${\tilde{d}_L}$  & 14&TeV\\
 ${\chi_3^0(\tilde H)}$        & 624&GeV     & ${\tilde{d}_R}$  & 13&TeV\\
 ${\chi_4^0(\tilde w)}$        & 1.5&TeV      & ${\tilde{t}_1}$  & 12&TeV   \\
 ${\chi_1^\pm}$        & 619&GeV      & ${\tilde{t}_2}$  & 13&TeV   \\
 ${\chi_2^\pm}$   & 1.5&TeV     & ${\tilde{b}_1}$  & 12&TeV  \\
 ${\tilde \ell_L}$   & 5.63&TeV      & ${\tilde{b}_2}$  & 13&TeV   \\
 \cline{4-6}
  \cline{4-6}
 ${\tilde e_R}$     & 2.8&TeV    & ${h}$  & 126.4&GeV  \\
 ${\tilde\tau_1}$     & 2.3&TeV      & ${H}$  & 2.2&TeV  \\
 ${\tilde\tau_2}$     & 5.5&TeV     & ${A}$  & 2.2&TeV    \\
 ${\tilde \n}$     & 5.5&TeV      & ${H^\pm}$  & 2.3&TeV   \\
\hline
\end{tabular}
\hspace{1cm}
\begin{tabular}{||l|l||} 
\hline
 $\tan\b$  & 55  \\
 sign($\mu$)  & $-1$  \\
 $N_M$ & $2$\\
 $M_{D} = M_{\bar D}$  & $1.965\times 10^6$\,GeV \\
 $M_{L}= M_{\bar L}$  & $1.225\times 10^6\,{\rm GeV}$ \\
 $k_D F/(M_DM_{\bar D})$  & $0.610$ \\
 $k_L F/(M_LM_{\bar L})$  & $0.999$ \\
 $m_t$(pole)  & $173.21$GeV \\
 $\alpha_s(m_Z) $ & $0.1185$\\
 \hline
\end{tabular}
\end{center}
\caption{A sample mass spectrum which explains the $Z$-boson signal
following Ref.~\cite{Lu:2015wwa}.
At this sample point, we take $k_{D,L}S_0 = M_{D,L}$, respectively.
We also require $B(M_{\rm mess}) \simeq 0$ by assuming minimal $\mu$-term,
as detailed in section \ref{sec:model}.
Due to the choice of $k_L F/(M_LM_{\bar L})\simeq 0.999$,
the lightest $L$-type messenger scalar particle is light, with a mass of around 27\,TeV.
We calculate the Higgs boson mass using {\tt SusyHD}~\cite{Vega:2015fna}
whose theoretical uncertainties from higher-order corrections are estimated to be about $1$\,GeV 
(see also discussions in section\,\ref{sec:gluino}).
}
\label{tab:spectrum}
\end{table}

To attain a light Higgsino, we remind readers that the $\mu$-term is related to $m_{H_u}^2$ via 
the electroweak symmetry breaking condition
\begin{eqnarray}
 \mu^2 \simeq - m_{H_u}^2(m_{\rm stop}) \simeq 
- m_{H_u}^2(M_{\rm mess}) + \frac{12  y_t^2} {16\pi^2} \,m_{\rm stop}^2 \,\log\frac{M_{\rm mess}}{m_{\rm stop}}\ .
\end{eqnarray}
Here we assume $\tan\b\gtrsim 50$, as discussed in the previous section.
This relation shows that a slightly larger $m_{H_u}^2(M_{\rm mess})$ for a given $m_{\rm stop}$ 
at the messenger scale results in a smaller $\mu$-term and hence lighter Higgsinos.
A larger $m_{H_u}^2$ also corresponds to larger bino and wino masses,
which are also favorable to explain the signal.
Through a careful parameter choice, we find that the desired spectrum 
can be achieved, as given in Table\,\ref{tab:spectrum}),
where the Higgsino masses are placed between the gluino mass and the bino mass.%
\footnote{We assume a somewhat heavy gravitino, $m_{3/2}\gtrsim 100$\,keV, 
so that the bino NLSP is stable inside the detector (see Eq.(\ref{eq:length})).}

At the model point in Table\,\ref{tab:spectrum}, the gluino decay is dominated by
the radiatively induced two-body modes with the branching ratios
\begin{eqnarray}
Br(\tilde g \to \tilde H + g )&\simeq &0.47\ ,\\
Br(\tilde g \to \tilde B + g )&\simeq& 0.06\ .
\end{eqnarray}
Here we use {\tt SDECAY\,v1.3}~\cite{Muhlleitner:2003vg} to calculate the decay widths of MSSM particles.
It should be noted that the MSSM parameters in Table\,\ref{tab:spectrum} is not optimal for the dominance 
of the two-body decay modes.  Thus, a rather light gluino is required to account for 
the observed signals\,\cite{Lu:2015wwa}. 
The production cross section for a pair of gluinos is given by
\begin{eqnarray}
\sigma = 132 \pm 13\, {\rm fb}\ ,
\end{eqnarray}
as calculated at the next-to-leading-logarithmic accuracy  
by {\tt NLL-Fast\,v1.2}\,\cite{Beenakker:1996ch,Beenakker:2015rna}.

For a simplified estimate, we rely on the analyses given in {\tt CheckMATE v1.2.1}\,\cite{Drees:2013wra}
which incorporates {\tt DELPHES\,3}\,\cite{deFavereau:2013fsa} and {\tt FastJet}\,\cite{Cacciari:2006sm}.
We generate signal events using {\tt MadGraph5  v2.2.3}\,\cite{Alwall:2011uj} connected to {\tt Pythia 6.4}\,\cite{Sjostrand:2006za}. 
The MLM matching scheme is used with a matching scale at $150$\,GeV~\cite{Alwall:2007fs} .
We choose  {\tt CTEQ6L1}\,\cite{Pumplin:2002vw} for the parton distribution functions.
As a result, we obtain about $10$ events in the signal region\,\cite{Aad:2015wqa},
while evading the constraints from the multi-jets$+E_{T}^{\rm miss}$ search\,\cite{Aad:2014wea},
the mono-jet search\,\cite{Aad:2015zva}, as well as the CMS on-$Z$ search~\cite{Khachatryan:2015lwa} at 95\%CL.%
\footnote{Here we do not take into account the constraints on the $ZZ$ mode from the 
four lepton $+E_T^{\rm miss}$ searches\,\cite{Chatrchyan:2014aea,Aad:2014iza}.
Due to smaller branching ratios of the gluino into a Higgsino and a gluon in our model,
the constraints from those searches are weaker than the one discussed in Ref.~\cite{Lu:2015wwa}.
We have also confirmed that the constraint from the $Z$+dijet$+E_{T}^{\rm miss}$ searches  
\cite{Aad:2014vma,Khachatryan:2014qwa} is less important.}
Therefore, the model parameters in Table\,\ref{tab:spectrum} can successfully provide signal events
consistent with the excess at $1.4\,\s$ level.%
\footnote{Here, we consider the $p$-value corresponding to the probability that the 
signal+background can explain the observed event numbers consistently,
and $1.4$\,$\s$ corresponds to $p \simeq 0.16$ (see {\it e.g.}, Ref.~\cite{Mistlberger:2012rs}).}

\section{Summary}

In this paper, we revisited a spacial model of gauge mediated supersymmetry breaking, the ``$R$-invariant direct gauge mediation.''
The model is favorable as it is durable even when the reheating temperature of the Universe is very high.
We paid particular attention  to the consistency of the model with the minimal model
addressing the origin of the $\mu$-term.
As a result, we found that the minimal model can be consistent with the $R$-invariant gauge mediation model
with a careful choice of model parameters, although incompatibility was highlighted in view of the current 
experimental constraints on superparticle masses and the observed Higgs boson mass.
We also found that the $\mu$-term was generically smaller than the stop mass,
which might ease the electroweak fine-tuning problem while explaining the observed 
Higgs boson mass with a heavy stop mass of ${\cal O}(10)$\,TeV.

We found that there existed an upper limit on the gluino mass from the observed Higgs boson mass
when the $\mu$-term was given by the minimal model.
Due to a hierarchy between gaugino masses and sfermion masses as well as the requirement 
for a large $\tan\b$, the observed Higgs boson mass led to an upper limit on the stop mass of about $20$\,TeV
and a corresponding upper limit on the gluino mass of about $4$\,TeV.
This result is encouraging because
the LHC experiment will be able to cover a large portion of the parameter space
unless the model parameters are highly optimized to achieve a large gluino mass.
This situation is parallel to, for example, high-scale supersymmetry breaking models with 
anomaly mediated gaugino mass \,\cite{AMSB,Harigaya:2014sfa} 
such as pure gravity mediation model/minimal split SUSY\,\cite{Ibe:2006de,PGMs, Split SUSY} 
(see also {\it e.g.} Ref.\,\cite{Wells:2004di}),
which also predicts that the gluino is within the reach of the future collider 
experiments\,\cite{PGM2,Yamanaka,Cirelli:2014dsa,Low:2014cba}, 
while explaining the observed Higgs boson mass with a large $m_{\rm stop}$.

We also discussed whether the $R$-invariant direct gauge mediation model could explain the
$3\sigma$ excess of the $Z+$jets$+E_T^{\rm miss}$ events  reported by the ATLAS Collaboration\,\cite{Aad:2015wqa}. 
With carefully chosen parameters, we found it possible to explain the excess and
the masses of Higgsinos were placed in between those of gluino and bino.%
\footnote{As another interesting feature of the  $R$-invariant direct gauge mediation model, it is often accompanied 
by a pseudo Nambu-Goldstone boson associated with $R$-symmetry breaking, the $R$-axion.
With the gluino mass range suggested by the $Z+$jets$+E_T^{\rm miss}$, it is also possible to search for the 
$R$-axion which can be produced via gluon-fusion at the LHC experiment\,\cite{Goh:2008xz}.  }

Finally, we comment on some ideas that provide the appropriate size of the $\mu$-term.
In our analysis, we have only discussed that the required size of the $\mu$-term for a successful
electroweak symmetry breaking is in the TeV range or smaller.
Since we assume that the $\mu$-term is consistent with the $R$-symmetry,
the smallness of the $\mu$-term requires some additional symmetry.
One popular idea is to generate the $\mu$-term from the breaking 
of a Peccei-Quinn symmetry\,\cite{Peccei:1977hh} via a dimension-5 operator\,\cite{Kim:1983dt}.%
\footnote{The scalar partner of the axion could cause some cosmological problem in the gauge mediation scenario,
although we do not go into details in this paper.}
As another possibility, we propose to make use of a $Z_2$ symmetry (we name here $10$-parity) under which 
only $H_d$ and the MSSM matter fields incorporated in the ${\bf 10}$ representation of $SU(5)$ GUT group
change their signs:
\begin{eqnarray}
(Q_L, \bar{U}_R,\bar{E}_R) \to 
- (Q_L, \bar{U}_R,\bar{E}_R) \ ,&&
\quad 
 (\bar{D}_R, L_L) \to  (\bar{D}_R, L_L)\ , \quad
 \bar N_R \to \bar N_R \ , \cr
 H_u \to H_u \ , &&\quad H_d \to - H_d\ .
\end{eqnarray}
In this case, the small  $\mu$-term can be explained by a tiny breaking of the $10$-parity.


\section*{Acknowledgments}

The authors thank S.\,Shirai for useful discussions on the realization of $Z+$jets$+E_T^{\rm miss}$ signal in SUSY models.
This work is supported in part by the Ministry of Science and Technology of Taiwan under Grant Nos.~MOST-100-2628-M-008-003-MY4 and 104-2628-M-008-004-MY4 (C.-W.~C); Grants-in-Aid for Scientific Research from the Ministry of Education, Culture, Sports, Science, and Technology (MEXT) KAKENHI, Japan, No. 24740151, No. 25105011 
and No. 15H05889 (M.~I.) as well as No. 26104009 (T.~T.~Y.); Grant-in-Aid No. 26287039 (M.~I. and T.~T.~Y.) from the Japan Society for the Promotion of Science (JSPS) KAKENHI; and by the World Premier International Research Center Initiative (WPI), MEXT, Japan (M.~I., and T.~T.~Y.).
K.H. was supported in part by a JSPS Research Fellowship for Young Scientists.
This work is also supported by MEXT Grant-in-Aid for Scientific research on Innovative Areas (No.15H05889).
%

\appendix

\section{$R$-charge assignments}\label{sec:symmetry}
\label{sec:Rcharge}

In this appendix, we summarize the $R$-charge assignments that allow the messengers to decay into the MSSM fields via a small mixing with the MSSM multiplet
\begin{eqnarray}
\label{eq:MESS-MATTER}
W \sim \vev{W} \Psi^\prime \, \bar{\bf 5}_{\rm MSSM}\ .
\end{eqnarray}
Hereafter, we use $SU(5)$ GUT representations for the MSSM matter fields: $\bar{\bf 5}_{\rm MSSM} = (\bar{D}_R, L_L)$ 
and ${\bf 10} = (Q_L, \bar{U}_R,\bar{E}_R)$.
By assuming that the messenger-SUSY breaking interactions given in Eq.\,(\ref{eq:SUSY-MESS}), 
the minimal $\mu$-term, the MSSM Yukawa interactions and the mass term of  
the right-handed neutrinos are consistent with the $R$-symmetry, we obtain the $R$-charge assignments given in Table\,\ref{tab:charge}. 
Here the charge assignments for the messenger fields are different from the one discussed in section\,\ref{sec:RDGMB},
which can be obtained by appropriately mixing the $R$-symmetry and messenger rotation.
It should be also noted that $\Psi^\prime$ in Eq.\,(\ref{eq:MESS-MATTER}) can be replaced with $\Psi$, leading
to different $R$-charge assignments (see Table\,\ref{tab:charge2}).

\begin{table}[t]
\caption{\sl\small
The $R$-charge assignments that allow the messenger-SUSY breaking interactions in Eq.\,(\ref{eq:SUSY-MESS}), 
the minimal $\mu$-term, the MSSM Yukawa interactions, the mass term of  
the right-handed neutrinos, and the messenger-matter mixing in Eq.\,(\ref{eq:MESS-MATTER}).
}
\begin{center}
\begin{tabular}{|c|c|c|c|c|c|c|c|c|c|c|}
\hline
& $S$ & ${\Psi} $ &
${\bar \Psi} $ &
${\Psi}^\prime $ &
 $\bar\Psi^\prime$ &
${H}_u$ & $H_d$ & ${\bf 10}_{\rm MSSM}$ & $\bar{\bf 5}_{\rm MSSM}$& $\bar{N}_R $

\\
\hline
$R$& $2$ & $11/5$ & $-11/5$ & $-1/5$ & $21/5$ & $4/5$& $6/5$& $3/5$& $1/5$ & $1$
\\
\hline
\end{tabular}
\end{center}
\label{tab:charge}
\end{table}%

Through the small mixing term, the lightest messenger decays into  MSSM particles with a decay width
\begin{eqnarray}
\G_{\rm mess} \sim \frac{g_a^2}{16\pi} \frac{m_{3/2}^2}{M_{\rm mess}}\ ,
\end{eqnarray}
which corresponds to the decay temperature
\begin{eqnarray}
 T_{\rm decay} \sim {\cal O}(1)\,{\rm GeV} \left(\frac{m_{3/2}}{10\,{\rm keV}}\right)
  \left(\frac{10^6\,\rm GeV}{M_{\rm mess}}\right)^{1/2}\ .
\end{eqnarray}
This decay temperature is much higher than the temperature at which the messenger field would 
dominate over the energy density of the Universe,
\begin{eqnarray}
 T_{\rm dom} \sim M_{\rm mess} Y_{\rm mess}\ ,
\end{eqnarray}
where the thermal yield of the lightest messenger (the doublet messenger)\,\cite{Fujii:2002fv} 
\begin{eqnarray}
Y_{\rm mess} \sim 10^{-10}\left( \frac{M_{\rm mess}}{10^{6}\,\rm GeV}\right) \ .
\end{eqnarray}

\begin{table}[t]
\caption{\sl\small
The $R$-charge assignments when $\Psi^\prime$ in Eq.\,(\ref{eq:MESS-MATTER}) is replaced by $\Psi$.
}
\begin{center}
\begin{tabular}{|c|c|c|c|c|c|c|c|c|c|c|}
\hline
& $S$ & ${\Psi} $ &
${\bar \Psi} $ &
${\Psi}^\prime $ &
 $\bar\Psi^\prime$ &
${H}_u$ & $H_d$ & ${\bf 10}_{\rm MSSM}$ & $\bar{\bf 5}_{\rm MSSM}$& $\bar{N}_R $

\\
\hline
$R$& $2$ & $-1/5$ & $1/5$ & $9/5$ & $11/5$ & $4/5$& $6/5$& $3/5$& $1/5$ & $1$
\\
\hline
\end{tabular}
\end{center}
\label{tab:charge2}
\end{table}%

\end{document}